\documentclass[10pt,twocolumn]{article}

\usepackage[letterpaper,margin=0.75in]{geometry}
\usepackage[T1]{fontenc}
\usepackage{newtxtext}
\usepackage{microtype}
\usepackage{graphicx}
\usepackage{caption}
\usepackage{natbib}
\usepackage[hyphens]{url}
\usepackage{xurl}
\usepackage{placeins}
\usepackage{dblfloatfix}
\usepackage{booktabs}
\usepackage{amsmath}
\usepackage{amssymb}
\usepackage[colorlinks=true,allcolors=blue]{hyperref}
\hypersetup{
  pdftitle={ReLMCodec: Designing Predictable Speech Tokens from Pre-Quantization Phoneme Structure},
  pdfauthor={Zixiang Wan, Xusheng Yang, Zheng Wang, Peiji Yang},
  pdfsubject={Low-bitrate neural speech codec with predictable single-stream tokens},
  pdfkeywords={speech codec, speech tokenization, autoregressive modeling, phoneme structure}
}

\newcommand{\cmark}{\checkmark}
\newcommand{\xmark}{\(\times\)}
\title{ReLMCodec: Designing Predictable Speech Tokens from Pre-Quantization Phoneme Structure}
\author{%
Zixiang Wan\textsuperscript{1,\ensuremath{\dagger}}\quad
Xusheng Yang\textsuperscript{1}\quad
Zheng Wang\textsuperscript{2}\quad
Peiji Yang\textsuperscript{2,*}\\[0.5em]
\small \textsuperscript{1}Peking University, Shenzhen, China\\
\small \textsuperscript{2}Tencent, Shenzhen, China\\[0.15em]
\small \href{mailto:zxwan25@stu.pku.edu.cn}{zxwan25@stu.pku.edu.cn}\quad
\href{mailto:peijiyang@tencent.com}{peijiyang@tencent.com}
}
\date{}

\begin{document}

\maketitle

\begingroup
\renewcommand{\thefootnote}{\fnsymbol{footnote}}
\footnotetext[1]{Corresponding author.}
\footnotetext[2]{This work was completed during an internship at Tencent.}
\endgroup

\begin{abstract}
Neural speech codecs face a fundamental tension in the language-model era: tokens that support high-fidelity reconstruction are not necessarily easy for autoregressive models to predict. Our controlled analysis of diverse codec and self-supervised speech representations shows that clearer phoneme structure before discrete code assignment is consistently associated with easier autoregressive token prediction. Yet phoneme structure alone is insufficient for high-fidelity reconstruction, which also requires reconstruction-relevant acoustic detail. Guided by this observation, we introduce ReLMCodec, a low-bitrate single-codebook speech codec built upon a preserve--control--refine principle: it preserves the linguistic organization of frozen self-supervised learning (SSL) features at the quantizer input, controls reconstruction-driven drift through Pre-quantization Anchor-Preserving Adaptation (PAPA), and refines the quantized latent space with a training-only WavLM-Large L24 teacher to reduce phoneme-level token fragmentation. Together, these components allow acoustic detail to support waveform reconstruction while keeping the resulting token sequence predictable for autoregressive models. At 650 and 800 bps, ReLMCodec moves the empirical single-stream predictability--reconstruction frontier in our evaluations, with gains that carry over to downstream text-to-speech (TTS) synthesis in both intelligibility and speaker similarity.
\end{abstract}

\noindent\textbf{Project page:} \url{https://github.com/ggiggit/ReLMCodec}

\section{Introduction}

As large language models extend to the speech generation domain, neural speech codecs are no longer only waveform compression modules. They have become the discrete interface between continuous speech and autoregressive language models~\citep{chen2024vall,ye2025llasa,defossez2024moshi}. Effective speech tokenization must therefore satisfy multiple requirements at once: it should support high-quality reconstruction, preserve stable linguistic organization, and form a token sequence with low next-token uncertainty.

Existing speech codecs broadly follow two architectural patterns. Residual vector quantization (RVQ)-based multi-codebook codecs progressively reduce quantization error and provide strong reconstruction, but their multiple interdependent token streams complicate autoregressive speech modeling~\citep{defossez2022high,kumar2023high,wan2026phoenixcodec,yang2025u}. Single-codebook codecs avoid this interface burden, but concentrate the semantic--acoustic trade-off in one bottleneck: reconstruction-driven training tends to allocate limited codebook capacity to speaker, prosodic, and local acoustic variation, whereas directly quantized SSL features often provide cleaner linguistic organization but lack reconstruction detail. As a result, \textit{tokens that are effective for waveform reconstruction are not necessarily the tokens that an autoregressive language model can predict most easily}~\citep{ye2025codec}.

Prior language-model-oriented codec designs suggest that adding semantic or SSL information can improve token predictability~\citep{zhang2024speechtokenizer,du2025cosyvoice,ye2025llasa}. However, it remains unclear which property of these representations is responsible for the improvement, and whether that property remains useful when the tokens must also support waveform reconstruction. This gap makes it difficult to design codec tokens that are both predictable for language models and effective for speech reconstruction.

To identify what makes codec tokens easier for autoregressive models to predict, we introduce a controlled probing protocol. We focus on the \emph{pre-quantization representation}, defined as the continuous frame-level representation that an encoder provides to the quantizer before discrete code assignment, and ask whether its \emph{phoneme structure}---how clearly speech frames are grouped by phoneme identity---predicts next-token modeling difficulty. In the probe, we freeze 24 codec and SSL representations, bypass native quantizers where present, and evaluate all representations with the same 8{,}192-codeword probing vector quantizer (P-VQ), the same autoregressive language model, and the same evaluation protocol. Under this matched setting, we find that representations with stronger phoneme structure before quantization produce higher next-token accuracy and lower perplexity under the probe. At the same time, phoneme structure alone does not explain reconstruction quality, motivating a codec design that preserves linguistic organization while adding reconstruction-relevant acoustic detail.

This finding leads to ReLMCodec's \textbf{preserve--control--refine} design. ReLMCodec \textbf{preserves} the linguistic organization of frozen W2v-BERT 2.0 features at the quantizer input~\citep{barrault2023seamless}, \textbf{controls} reconstruction-driven drift by adding acoustic detail through Pre-quantization Anchor-Preserving Adaptation (PAPA), and \textbf{refines} the quantized latent space with a training-only WavLM-Large L24 teacher so that reconstruction training does not unnecessarily split phoneme-related structure across tokens~\citep{chen2022wavlm}. Together, these components allow acoustic detail to support waveform reconstruction while keeping the resulting token sequence predictable for autoregressive models.

Our contributions are fourfold:
\begin{itemize}
    \item \textbf{Controlled diagnosis.} A matched probing protocol over 24 codec and SSL representations isolates quantizer-input structure under fixed discrete capacity and language-model optimization without a waveform decoder. Phoneme separability strongly tracks predictability, with k-nearest-neighbor (KNN) accuracy associated with probe accuracy and perplexity and supported by cluster metrics.
    \item \textbf{Design principle.} The analysis shows that predictable tokens are associated with stable linguistic organization before quantization, while high-quality reconstruction also requires acoustic detail. This motivates a preserve--control--refine principle for coordinating these two requirements around the quantizer input.
    \item \textbf{Technical mechanism.} We instantiate this principle with PAPA, a capacity-matched reparameterization that preserves frozen SSL features as an explicit anchor and adds acoustic detail through a fixed-scale residual path, without adding parameters relative to direct adaptation.
    \item \textbf{End-to-end validation.} ReLMCodec advances the empirical low-bitrate single-codebook frontier, achieving leading reconstruction quality among evaluated checkpoints while showing corresponding gains in downstream TTS intelligibility and speaker similarity.
\end{itemize}

\section{Related Work}

\subsection{Low-Bitrate Single-Stream Neural Speech Codecs}

Low-bitrate single-stream codecs reduce the sequence burden for speech language modeling. WavTokenizer, BigCodec, and FocalCodec improve low-rate reconstruction through compact bottlenecks and strong acoustic modeling~\citep{ji2025wavtokenizer,xin2024bigcodec,della2026focalcodec}; UniCodec, SemantiCodec, and LSCodec extend this direction through multi-domain modeling, semantic--acoustic encoding, or reduced speaker leakage~\citep{jiang2025unicodec,liu2024semanticodec,guo2025lscodec}. ReLMCodec targets a distinct interface property: a compact reconstructive stream may still be difficult for an autoregressive model to predict.

\subsection{Structured Semantic--Acoustic Speech Representations}

SpeechTokenizer, XY-Tokenizer, ContextCodec, and OmniCodec introduce linguistic structure through teacher guidance, text alignment, branch separation, or hierarchical organization~\citep{zhang2024speechtokenizer,gong2026xy,liang2026contextcodec,hu2026omnicodec}. ReLMCodec instead retains a single-codebook interface and parameterizes acoustic adaptation around an explicit frozen SSL anchor, followed by training-only post-quantization refinement.

\subsection{Unified and Autoregressively Predictable Speech Tokens}

Language-model-oriented tokenizers combine SSL--acoustic fusion, temporal compression, semantic-prior quantization, distillation, or predictive objectives~\citep{ye2025codec,ye2025llasa,yang2025almtokenizer,della2026wavslm,yang2024uniaudio}. Representation and layer selection are task dependent~\citep{mousavi2024should,wan2025metadata}, while PINT learns invariant content tokens without waveform reconstruction~\citep{wagner2026content}. ReLMCodec first diagnoses predictability under a matched quantizer and language model, then applies the resulting principle to an end-to-end single-codebook codec. Phoneme labels are used only for analysis.

\section{Pre-quantization Structure Predicts Token Predictability}

This section follows four linked steps. We first construct a controlled diagnostic that fixes quantizer capacity and language-model optimization across representations. We then test whether phoneme structure before quantization predicts token modeling difficulty, examine whether that structure remains visible after discrete assignment, and finally translate the diagnosis into separate representation roles for codec reconstruction and training-time refinement. This progression separates three levels of evidence: continuous quantizer-input representations, discrete token assignments, and the codec design that follows from them.

\subsection{Controlled Diagnostic Protocol}

\noindent\textbf{Datasets.}
We use LibriSpeech~\citep{panayotov2015librispeech} with phoneme boundaries generated by the Montreal Forced Aligner (MFA)~\citep{mcauliffe2017montreal}. A fixed-seed sample of 500 train-960 and 100 test utterances defines only the phoneme-analysis sets; token language-model probing uses train-960 and the test splits identically across representations. Appendix A specifies the alignment configuration and frame-label construction.

\noindent\textbf{Pipeline.}
We bypass native quantizers, standardize frozen features, and attach P-VQ: a single-codebook probe with exponential moving average (EMA) updates, 8{,}192 (8K) eight-dimensional codewords, trainable projections, and no waveform decoder. It is trained for 200K steps with feature-reconstruction and commitment objectives. For each representation, we separately train a Qwen2-1.5B model on its resulting token sequences, using identical parameter initialization, dataset splits, and optimization settings across all models~\citep{yang2024qwen2}; Appendix B reports the full probe and token language-model configurations.

\noindent\textbf{Representations.}
We evaluate 24 codec/tokenizer and SSL representations under the common probing protocol. Appendix C and Appendix Table~A3 give the complete baseline set, selected layers, and checkpoint sources.

\noindent\textbf{Primary metrics.}
We operationalize phoneme structure with KNN phoneme accuracy; Silhouette, Davies--Bouldin, and V-measure provide complementary checks. Probe-ACC (P-ACC) and Probe-PPL (P-PPL) respectively denote the top-1 next-token prediction accuracy and perplexity obtained by evaluating the autoregressive language model on P-VQ token sequences. Correlations use all $n=24$ evaluated 50-Hz representations and exclude ReLMCodec.

\subsection{Correlation with Token Modeling Difficulty}

The first question is whether phoneme structure at the quantizer input is associated with the difficulty of modeling the resulting token sequence. Table~\ref{tab:correlations} tests this relationship across all 24 representations under the matched probing protocol.

\begin{table}[!ht]
\centering
\small
\setlength{\tabcolsep}{1pt}
\begin{tabular}{@{}lcc@{}}
\toprule
Metric & $\rho$ vs. P-ACC & $\rho$ vs. P-PPL \\
\midrule
KNN $\uparrow$ & 0.911 [0.76, 0.98] & -0.901 [-0.98, -0.73] \\
Silhouette $\uparrow$ & 0.715 [0.39, 0.88] & -0.713 [-0.88, -0.40] \\
Davies--Bouldin $\downarrow$ & -0.837 [-0.93, -0.61] & 0.848 [0.63, 0.94] \\
V-measure $\uparrow$ & 0.852 [0.67, 0.93] & -0.857 [-0.94, -0.68] \\
\bottomrule
\end{tabular}
\caption{Spearman correlations between phoneme separability and predictability of P-VQ tokens over the 24 evaluated representations. Brackets give 95\% bootstrap confidence intervals.}
\label{tab:correlations}
\end{table}

Across all 24 representations, KNN has the strongest rank association with both P-ACC and P-PPL. V-measure and Davies--Bouldin show the same pattern, whereas Silhouette is weaker, suggesting that local phoneme consistency is more informative than global cluster compactness alone. All results use 10{,}000 bootstrap resamples and have two-sided permutation $p<0.001$. Thus, under fixed quantization and language-model optimization, stronger pre-quantization phoneme structure is associated with easier token modeling. This is an observational statement and does not imply that phoneme structure alone determines reconstruction quality. Appendix Table~A4 reports the complete statistics.

This establishes the first link in the analysis: the phoneme structure of the continuous representation entering the quantizer is predictive of the difficulty of modeling the resulting tokens. The next question is whether this structure remains visible after the representation is discretized.

\subsection{Structure Transfer from Representation to Token Assignment}

\begin{figure}[t]
\centering
\includegraphics[width=0.49\columnwidth]{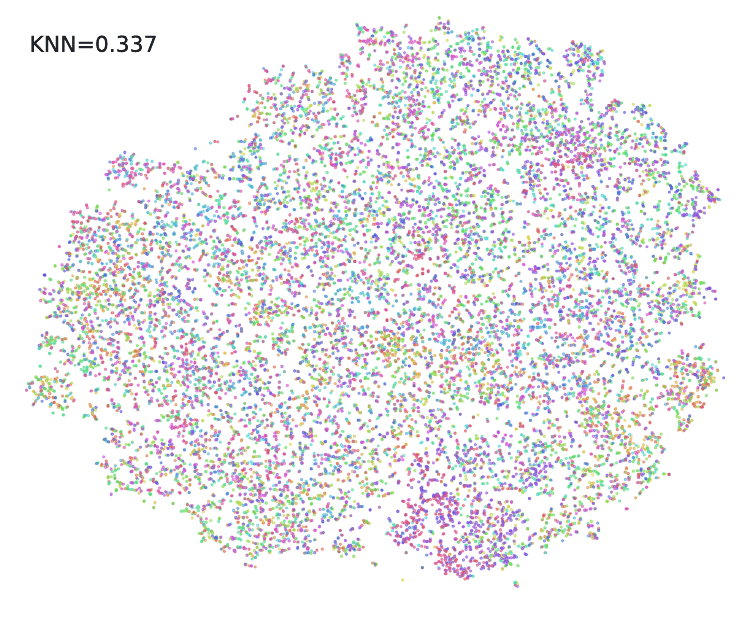}\hfill
\includegraphics[width=0.49\columnwidth]{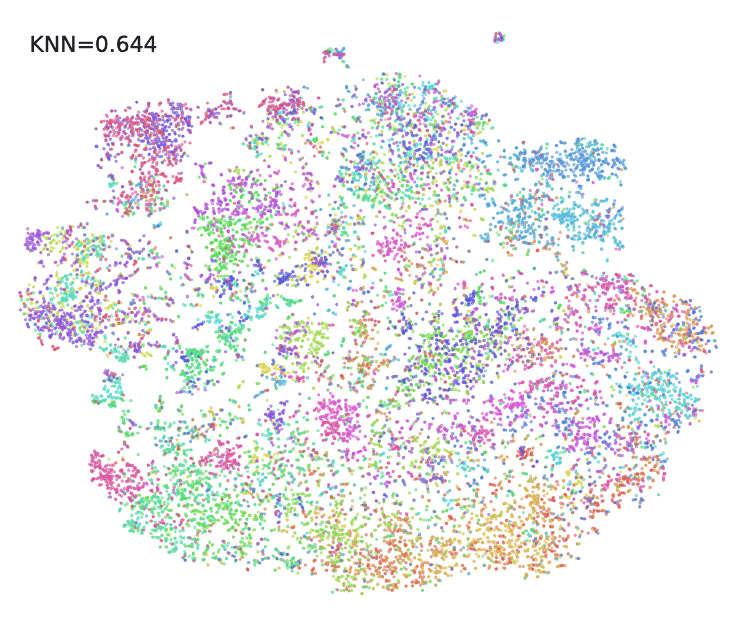}
\includegraphics[width=\columnwidth]{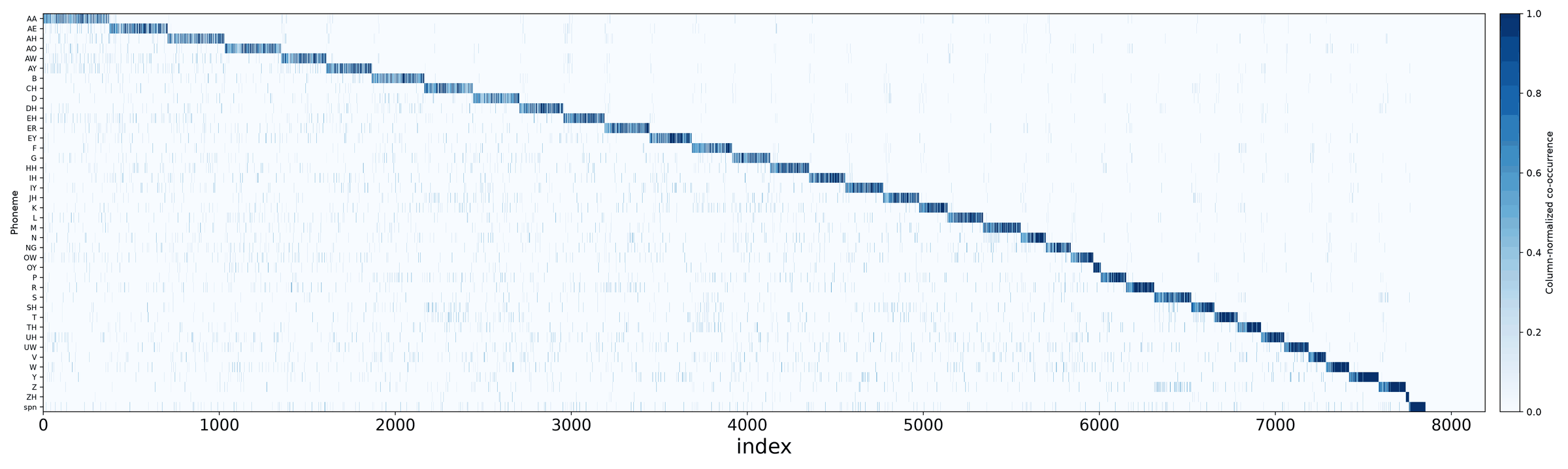}\par
\includegraphics[width=\columnwidth]{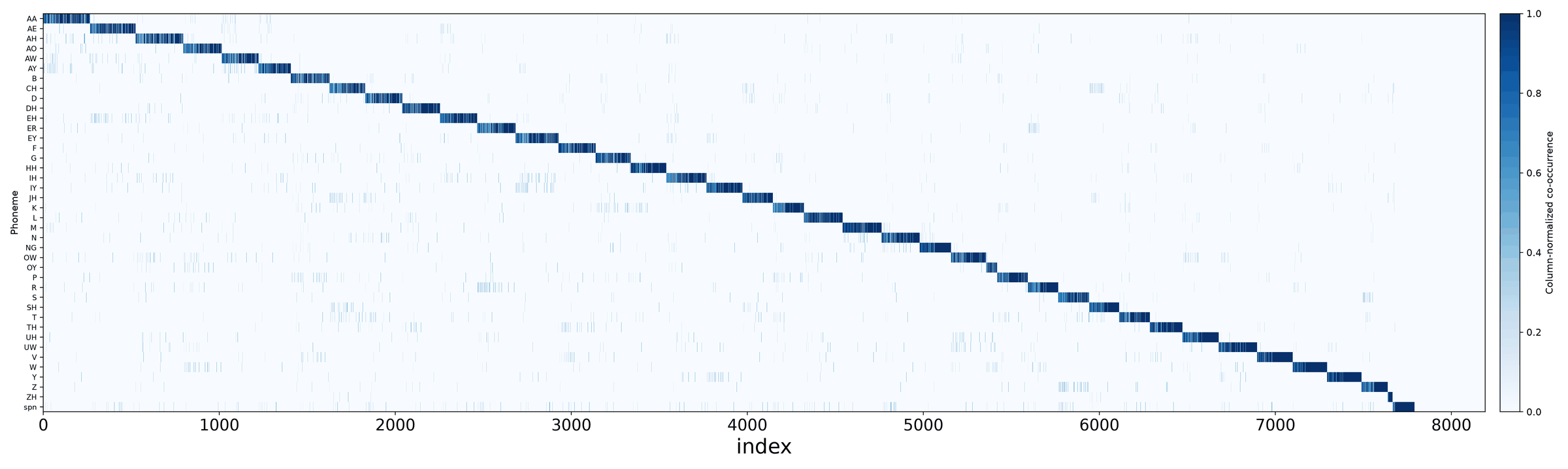}
\caption{Structure transfer from pre-quantization representations to matched P-VQ assignments. Top: frame-level representations colored by phoneme labels. Middle and bottom: token--phoneme co-occurrence after matched 8K P-VQ assignment. Concentrated blocks indicate that phoneme structure remains visible after discrete assignment. Additional examples are provided in Appendix Figures~A2--A4.}
\label{fig:representation-geometry}
\end{figure}

The second question is whether the phoneme structure measured before quantization is still reflected in the discrete assignments produced by the matched P-VQ. Figure~\ref{fig:representation-geometry} traces this process from the pre-quantization representation to token--phoneme co-occurrence. Appendix Figures~A2--A4 provide additional examples.

\FloatBarrier

For the pre-quantization view, Figure~\ref{fig:representation-geometry} shows frame-level representations colored by phoneme labels. Clearer phoneme structure appears as more coherent phoneme-conditioned regions. For the post-quantization view, we assign the same MFA-aligned, phoneme-balanced frames using the separately trained 8K P-VQ for each representation. Each heatmap column corresponds to a token and is normalized across phonemes; token columns are grouped by their dominant phoneme. Concentrated blocks indicate that the discrete tokens remain more consistently associated with specific phoneme classes, whereas broader off-block mass indicates stronger sharing of tokens across phonemes.

Under this trace, clearer phoneme structure before quantization is accompanied by sharper token--phoneme co-occurrence after quantization. This suggests that the structure measured at the quantizer input is not only correlated with token predictability, but can also remain expressed in the matched discrete assignments. Together with Table~\ref{tab:correlations}, Figure~\ref{fig:representation-geometry} supports the intended diagnostic chain: phoneme structure is measured before quantization, is associated with easier token modeling, and remains visible after discrete assignment.

\subsection{Design Implication: Separate Main and Teacher Roles}

The analysis leaves a design question: which representation should carry this structure in a waveform codec, and which representation should guide refinement? A codec representation cannot be selected only for phoneme structure, because it must also retain information needed for waveform reconstruction. A training-only teacher, by contrast, can prioritize stronger phoneme structure without serving as the reconstruction representation itself.

\begin{figure}[t]
\centering
\includegraphics[width=\columnwidth]{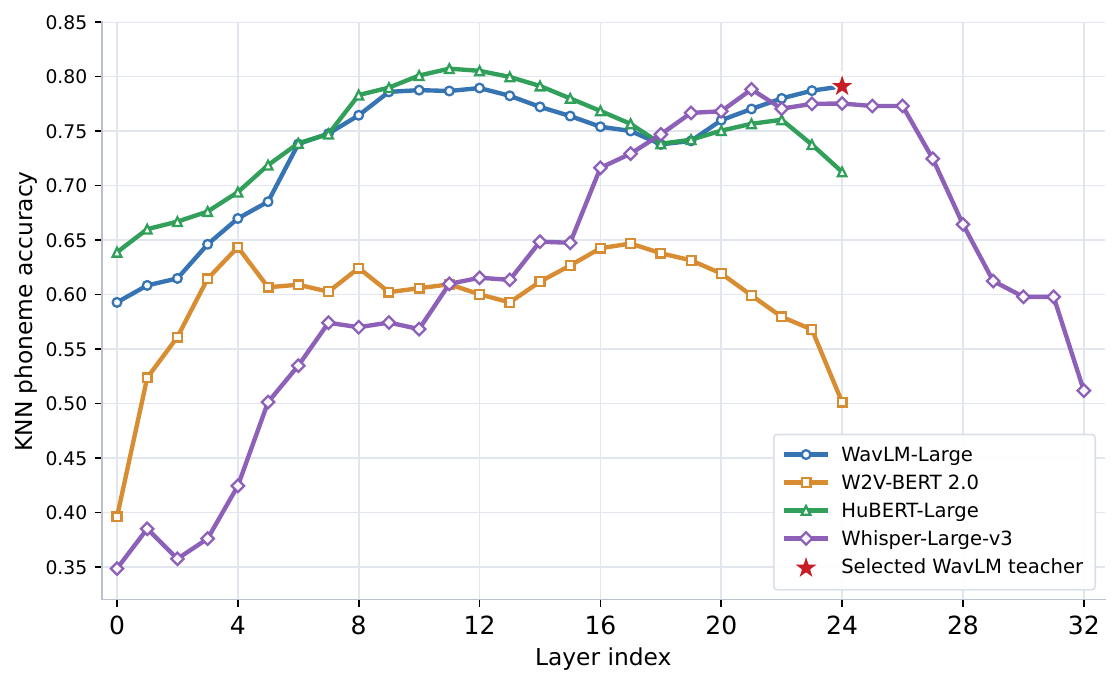}
\caption{Layer-wise KNN phoneme accuracy for four encoders; the star marks the selected WavLM-Large L24 teacher.}
\label{fig:layerwise-separability}
\end{figure}

Figure~\ref{fig:layerwise-separability} shows that phoneme structure is layer dependent and non-monotonic. HuBERT peaks in the middle layers, Whisper rises late and then drops sharply, and WavLM maintains a broad high-separability region through L24. The design therefore does not treat an encoder's final layer, or even its maximum-KNN layer, as universally optimal. W2v-BERT 2.0 L17 is retained as the reconstruction-compatible main path, while WavLM-Large L24 provides strong phoneme structure and the highest P-ACC among the evaluated SSL representations.

These different profiles motivate separating the reconstruction representation from a structure-oriented teacher rather than assigning the same SSL layer to both roles. The following section instantiates this principle in a codec architecture; the corresponding role ablation is reported with the other component studies.
\enlargethispage{\baselineskip}

\section{Method}

\subsection{Overall Architecture}

\begin{figure*}[t]
\centering
\includegraphics[width=\textwidth]{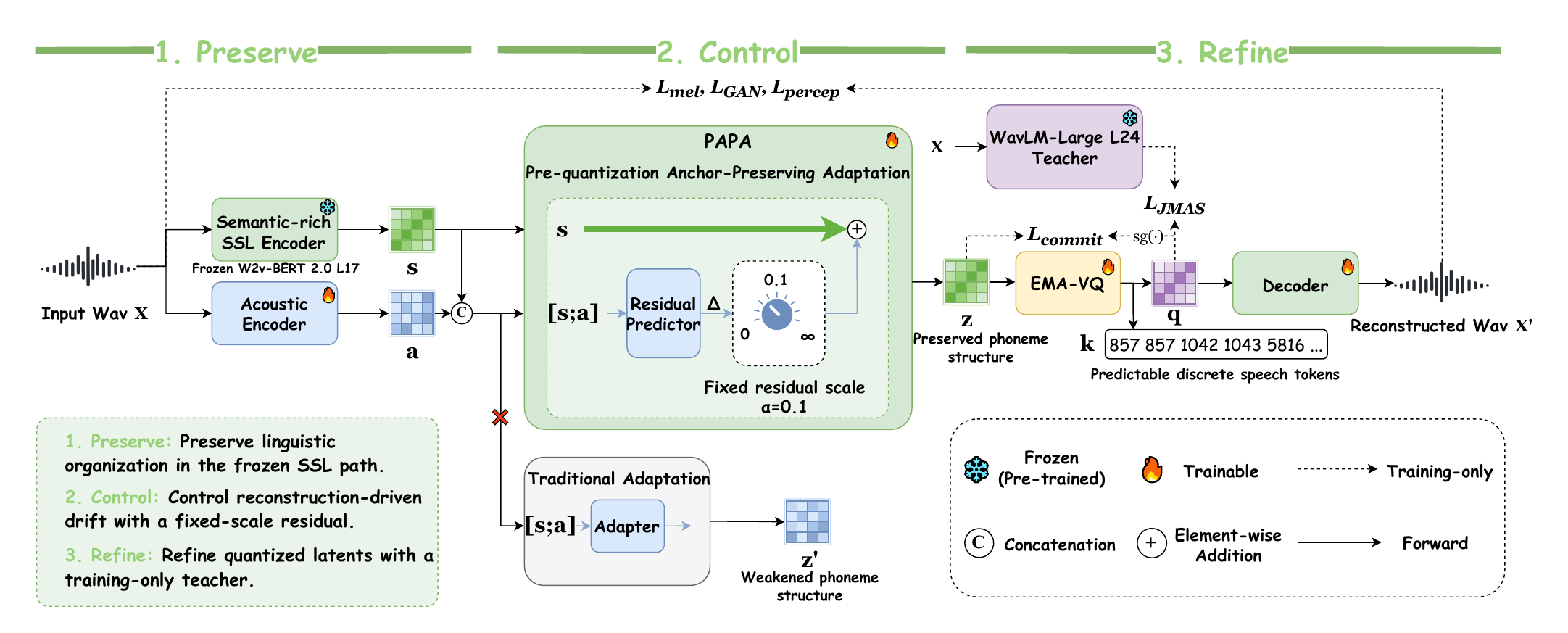}
\caption{Overall architecture of ReLMCodec.}
\label{fig:architecture}
\end{figure*}

ReLMCodec follows a preserve--control--refine design for single-stream speech tokenization. 
\emph{Preserve} keeps a frozen SSL path at the quantizer input to retain phoneme-structured information. 
\emph{Control} parameterizes reconstruction-relevant acoustic detail as a fixed-scale residual, biasing adaptation away from unrestricted replacement of the SSL geometry. 
\emph{Refine} uses a training-only teacher to regularize the quantized latent representation toward teacher-aligned phoneme structure. 
As shown in Figure~\ref{fig:architecture}, ReLMCodec combines frozen W2v-BERT 2.0 L17 features, a trainable acoustic encoder, a 12-layer Pre-quantization Anchor-Preserving Adaptation (PAPA) predictor, a single-layer EMA vector quantizer (EMA-VQ), and a waveform decoder.

\subsection{Pre-quantization Anchor-Preserving Adaptation}

PAPA anchors SSL--acoustic adaptation at the quantizer input by writing the adapted representation as a frozen SSL feature plus a scaled residual correction. A direct adapter learns the entire quantizer input from fused SSL and acoustic features, which can overwrite the original SSL geometry. PAPA keeps the frozen SSL sequence as a coefficient-one term and lets the trainable branch predict only an additive correction. For each frame $t$, let the aligned 50-Hz features from the frozen SSL encoder and trainable acoustic encoder be
\begin{equation}
s_t=[E_{\mathrm{ssl}}(x)]_t, \qquad
a_t=[E_{\mathrm{ac}}(x)]_t, \quad t=1,\ldots,T.
\end{equation}
Both have 1024 channels. Their channel-wise concatenation is processed by a 12-layer residual predictor $R$, which maps the 2048-dimensional joint sequence back to the main-path dimension:
\begin{equation}
\Delta_t=[R([s;a])]_t, \qquad
z_t=s_t+\alpha\Delta_t.
\end{equation}
The resulting $z_t$ is used as the quantizer input at frame $t$.

\paragraph{Forward anchoring.}
The coefficient-one SSL term keeps each frozen feature $s_t$ explicitly present at the quantizer input, while the residual branch supplies reconstruction- and quantization-relevant corrections. This parameterization does not guarantee invariant distances or neighborhoods, since the learned residual can still change the geometry. Its role is to make any departure from the SSL representation explicit through the residual term.

\paragraph{Fixed-scale residual parameterization.}
We set $\alpha=0.1$ so the residual path starts local relative to the frozen SSL anchor. This scale is not a hard bound on residual magnitude; the predictor could learn a larger $\Delta_t$ and offset it. Its effect is a parameterization and optimization bias: under standard initialization, $z_t$ starts close to $s_t$, and the raw gradient entering the residual branch is scaled before adaptive normalization,
\begin{equation}
\nabla_{\theta_R}\mathcal{L}=\alpha J_R^{\top}\nabla_z\mathcal{L}.
\end{equation}
With AdamW, this factor is not a proportional bound on the final parameter update. We empirically measure the residual-to-anchor ratio in Table~\ref{tab:geometry-retention}.

PAPA remains quantization-aware: the acoustic encoder and residual predictor are optimized jointly through the commitment, reconstruction, adversarial, and post-quantization alignment objectives. It changes the parameterization rather than model capacity: matched direct-adaptation and residual-adaptation variants use the same predictor architecture, parameter count, initialization, optimizer, and schedule. Any suitable frozen encoder can provide the anchor after adjusting interface dimensions.

\subsection{Single-Codebook Vector Quantization}

Our single-layer EMA-VQ uses 8{,}192 (8K) or 65{,}536 (64K) codewords at 50 tokens per second (TPS), yielding 650/800 bits per second (BPS). For utilization, we use 8-dimensional factorized codes with $\ell_2$ normalization~\citep{kumar2023high}, code expiration, and a cross-batch feature pool~\citep{wan2025spectokenizer}. Training uses EMA updates and commitment loss.

\subsection{Post-quantization Structure Refinement}

We adapt the joint--marginal alignment loss from JMAS-VAE~\citep{cheng2026jmas} and refer to the resulting post-quantization regularizer as the JMAS loss. At frame $t$, let $y_t=[E_{\mathrm{WavLM\text{-}Large\text{-}L24}}(x)]_t$ be the frozen teacher feature, $k_t$ the discrete assignment, and $\tilde{z}_t\equiv q_t$ the selected codebook embedding propagated through the straight-through estimator. After output projection, $\tilde{z}_t$ and $y_t$ are frame-aligned 1024-dimensional representations at 50 Hz. JMAS acts on $\tilde{z}_t$ rather than directly on $k_t$:
\begin{equation}
\mathcal{L}_{\mathrm{JMAS}}
=\lambda_{\mathrm{frame}}\mathcal{L}_{\mathrm{frame}}
+\lambda_{\mathrm{struct}}\mathcal{L}_{\mathrm{struct}}.
\end{equation}

The frame term aligns teacher and quantized representations at each position:
\begin{equation}
\mathcal{L}_{\mathrm{frame}}
=\frac{1}{T}\sum_{t=1}^{T}
\operatorname{ReLU}\!\left(1-m_1-\cos(\tilde{z}_t,y_t)\right).
\end{equation}
The structure term aligns their within-sequence pairwise relations:
\begin{equation}
\begin{aligned}
\mathcal{L}_{\mathrm{struct}}
=\frac{1}{T(T-1)}\sum_{t\ne u}
\operatorname{ReLU}\!\big(&\lvert\cos(\tilde{z}_t,\tilde{z}_u)\\
&-\cos(y_t,y_u)\rvert-m_2\big).
\end{aligned}
\end{equation}
The two terms transfer local phonetic information and preserve relative similarity and dissimilarity patterns. We set $m_1=0.5$, $m_2=0.25$, and $\lambda_{\mathrm{frame}}=\lambda_{\mathrm{struct}}=1$, uniformly sampling at most 256 frames for the pairwise term. JMAS thereby regularizes the quantized latent representation and indirectly shapes the discrete tokens.

\subsection{Reconstruction and Training Objective}

The quantized representation is decoded by a 12-layer VocosBackbone with hidden size 1024, feed-forward network (FFN) dimension 4096, and an inverse short-time Fourier transform (ISTFT) head ($n_{\mathrm{fft}}=1280$, hop 320)~\citep{siuzdak2023vocos,ji2025wavtokenizer}. Adversarial training uses a multi-period discriminator with periods $[2,3,5,7,11]$~\citep{kong2020hifi} and a multi-scale short-time Fourier transform (STFT) discriminator with FFT sizes $\{78,126,206,334,542,876,1418,2296\}$~\citep{parker2025scaling}. The base and continuation objectives are
\begin{equation}
\begin{aligned}
\mathcal{L}_{\mathrm{base}}
&=45\mathcal{L}_{\mathrm{mel}}+1000\mathcal{L}_{\mathrm{commit}}
+45\mathcal{L}_{\mathrm{JMAS}}+\mathcal{L}_{\mathrm{GAN}},\\
\mathcal{L}_{\mathrm{cont}}
&=\mathcal{L}_{\mathrm{base}}+450\mathcal{L}_{\mathrm{perc}}.
\end{aligned}
\end{equation}
The generative adversarial network (GAN) loss is warmed up for 50K steps, and $\mathcal{L}_{\mathrm{perc}}$ is the multi-layer frame-wise normalized-$\ell_1$ loss from a frozen WavLM speaker-verification (WavLM-SV) model. W2v-BERT 2.0 L17, WavLM-Large L24, and WavLM-SV remain frozen. At inference, only W2v-BERT 2.0, the acoustic encoder, PAPA, the quantizer, and the waveform decoder are retained, so both WavLM networks add no inference-time computation.

\begin{table*}[t]
\centering
\small
\setlength{\tabcolsep}{5pt}
\begin{tabular}{@{}lccccccccc@{}}
\toprule
Codec & TPS & BPS & Codebooks & WER$\downarrow$ & SIM$\uparrow$ & Log-Mel$\downarrow$ & PESQ$\uparrow$ & STOI$\uparrow$ & UTMOS$\uparrow$ \\
\midrule
Ground Truth & -- & -- & -- & 3.63 & 1.000 & 0.000 & 4.64 & 1.000 & 3.78 \\
DAC & 50 & 1000 & 2 & 27.82 & 0.310 & 2.110 & 1.14 & 0.730 & 1.30 \\
SpeechTokenizer & 50 & 1000 & 2 & 10.45 & 0.339 & 1.993 & 1.24 & 0.757 & 2.10 \\
X-Codec & 50 & 1000 & 2 & 5.26 & 0.680 & 1.473 & 2.13 & 0.889 & 3.89 \\
Stable Codec & 25 & 700 & 2 & 11.18 & 0.573 & 2.004 & 2.08 & 0.887 & \underline{4.06} \\
XY-Tokenizer & 12.5 & 1000 & 8 & 4.79 & 0.795 & 1.381 & 2.23 & 0.896 & 3.70 \\
mimi & 12.5 & 1100 & 8 & 7.71 & 0.723 & 1.838 & 2.18 & 0.892 & 3.33 \\
Qwen3-TTS-Tokenizer & 12.5 & 1100 & 8 & 6.93 & 0.589 & 1.842 & 1.35 & 0.846 & 2.56 \\
UniCodec & 75 & 1050 & 1 & 8.80 & 0.758 & 1.347 & \underline{2.38} & \underline{0.906} & 3.76 \\
WavTokenizer & 75 & 900 & 1 & 10.77 & 0.683 & \underline{1.330} & 2.30 & 0.899 & 3.72 \\
X-Codec2 & 50 & 800 & 1 & 5.53 & 0.801 & 1.392 & 2.13 & 0.884 & 3.76 \\
AUV & 50 & 716 & 1 & 7.07 & \underline{0.803} & 1.370 & 2.32 & 0.901 & 3.78 \\
SemantiCodec & 50 & 650 & 1 & 12.23 & 0.605 & 1.610 & 1.74 & 0.840 & 2.60 \\
FocalCodec & 50 & 650 & 1 & 5.01 & 0.749 & 1.776 & 1.40 & 0.848 & 3.86 \\
ReLMCodec@8K & 50 & 650 & 1 & \underline{4.16} & 0.749 & 1.370 & 2.17 & 0.900 & 4.03 \\
ReLMCodec@64K & 50 & 800 & 1 & \textbf{3.96} & \textbf{0.804} & \textbf{1.270} & \textbf{2.40} & \textbf{0.917} & \textbf{4.07} \\
\bottomrule
\end{tabular}
\caption{End-to-end reconstruction; @8K/@64K use the same architecture with 8{,}192/65{,}536 codewords.}
\label{tab:reconstruction}
\end{table*}

\section{Experiments}

\subsection{Experimental Setup}

\noindent\textbf{Datasets.}
ReLMCodec, P-VQ probes, token language models, and downstream TTS models are trained on LibriSpeech train-960~\citep{panayotov2015librispeech}. Speech reconstruction is evaluated on test-clean and test-other. For downstream TTS, we use EmoVoice~\citep{yang2025emovoice} as the shared architecture and training framework for all codec tokenizations, and evaluate on the fixed LibriSpeech subset from the F5-TTS protocol~\citep{chen2025f5}. All waveform outputs are resampled to 16 kHz before evaluation.

\noindent\textbf{Training.}
ReLMCodec is trained on 8 NVIDIA H20 96GB GPUs with AdamW, using a learning rate of $2\times10^{-4}$, $\beta_1=0.8$, $\beta_2=0.9$, 1000 warmup steps, and cosine decay. We use 3-second crops, a per-GPU batch size of 16, 32-bit floating-point (FP32) training, and gradient clipping at 1.0. Each ReLMCodec variant is trained for 200K base steps and then continued for another 200K steps with the frozen WavLM-SV perceptual loss.

\noindent\textbf{End-to-end codec comparison.}
Table~\ref{tab:reconstruction} compares released end-to-end baseline checkpoints without component retraining or replacement against our trained ReLMCodec checkpoints. All systems share metric implementations; Appendices C and E detail checkpoint and reconstruction protocols and computational cost and inference efficiency, respectively.

\noindent\textbf{Metrics.}
For speech reconstruction, we report word error rate (WER) computed with Whisper-Large-v3~\citep{radford2022whisper}, multi-scale log-mel spectrogram loss, short-time objective intelligibility (STOI), perceptual evaluation of speech quality (PESQ), speaker similarity (SIM) computed with WavLM~\citep{chen2022wavlm}, and model-based UTMOS~\citep{saeki2022utmos}. For downstream TTS, we report WER, SIM, and UTMOS.

The experiments ask three questions: does ReLMCodec improve the low-bitrate predictability--reconstruction operating point; which SSL role assignments and modules contribute; and do the resulting tokens improve downstream TTS?

\subsection{End-to-End Reconstruction and Predictability}

Table~\ref{tab:reconstruction} supports two rate-matched conclusions. At 650 bps, ReLMCodec@8K improves every metric over FocalCodec and SemantiCodec except matching FocalCodec's SIM, showing that the SSL anchor does not preclude competitive reconstruction. At 800 bps, ReLMCodec@64K improves WER/SIM/PESQ over X-Codec2 from 5.53/0.801/2.13 to 3.96/0.804/2.40, while the matched P-VQ probe raises P-ACC from 5.12\% to 9.65\%. ReLMCodec@64K leads all codec rows in WER, SIM, log-mel loss, PESQ, STOI, and UTMOS. The simultaneous same-rate gains indicate an improved predictability--reconstruction operating point. Appendix Figure A5 further places ReLMCodec@64K closest to ground truth in human mean opinion score for overall quality (H-MOS).

\subsection{Disentangling Preservation, Adaptation, and Refinement}

Table~\ref{tab:ssl-role-ablation} first tests SSL assignment to the reconstruction path and training-only teacher; KNN is measured at the quantizer input.

\begin{table}[!ht]
\centering
\small
\setlength{\tabcolsep}{1.5pt}
\begin{tabular}{@{}llccccc@{}}
\toprule
Main & Teacher & KNN & P-ACC & WER & SIM & PESQ \\
\midrule
W2B-L17 & W2B-L17 & 0.6347 & 9.31 & 4.65 & 0.747 & 2.21 \\
W2B-L17 & WLM-L24 & 0.6439 & 9.63 & 4.16 & 0.749 & 2.17 \\
WLM-L24 & WLM-L24 & 0.7905 & 28.59 & 5.58 & 0.738 & 1.74 \\
WLM-L24 & W2B-L17 & 0.7486 & 25.14 & 5.15 & 0.740 & 1.86 \\
\bottomrule
\end{tabular}
\caption{SSL-role ablation at 8K; W2B and WLM abbreviate W2v-BERT 2.0 and WavLM-Large L24, and Main/Teacher specify the PAPA anchor and structure target.}
\label{tab:ssl-role-ablation}
\end{table}

With the W2v-BERT 2.0 L17 main path fixed, replacing the W2v-BERT teacher with WavLM-Large L24 raises P-ACC from 9.31\% to 9.63\% and lowers WER from 4.65 to 4.16, with nearly unchanged SIM and a small PESQ decrease. Using WavLM-Large L24 as the main path instead yields much higher KNN and P-ACC but worse reconstruction metrics. The ablation therefore supports the asymmetric assignment without implying a universal teacher ranking.

With the asymmetric roles fixed, Table~\ref{tab:ablation} isolates SSL-only (A), acoustic-only (B), and combined (C) paths. In A/C, direct adaptation uses $z_t=R(u_t)$ and PAPA uses $z_t=s_t+\alpha R(u_t)$; paired configurations otherwise share architecture, initialization, optimizer, and schedule.

\begin{table}[!ht]
\centering
\small
\setlength{\tabcolsep}{0.9pt}
\begin{tabular}{@{}ccccccccccc@{}}
\toprule
ID & S & A & $R$ & PAPA & JMAS & P-ACC & WER & SIM & PESQ & UTMOS \\
\midrule
A0 & \cmark{} & \xmark{} & \xmark{} & \xmark{} & \xmark{} & \textbf{12.45} & 8.31 & .613 & 1.79 & 3.61 \\
A1 & \cmark{} & \xmark{} & \cmark{} & \xmark{} & \xmark{} & 3.12 & 6.68 & \textbf{.790} & \textbf{2.15} & 3.74 \\
A2 & \cmark{} & \xmark{} & \cmark{} & \cmark{} & \xmark{} & 10.51 & \underline{6.32} & \underline{.782} & \underline{2.13} & 3.70 \\
A3 & \cmark{} & \xmark{} & \cmark{} & \xmark{} & \cmark{} & 4.84 & 6.35 & .744 & 2.03 & \underline{3.78} \\
A4 & \cmark{} & \xmark{} & \cmark{} & \cmark{} & \cmark{} & \underline{10.70} & \textbf{6.21} & .742 & 2.03 & \textbf{3.79} \\
\midrule
B0 & \xmark{} & \cmark{} & \xmark{} & \xmark{} & \xmark{} & \underline{4.60} & \underline{5.27} & \underline{.592} & \underline{1.99} & \underline{3.83} \\
B1 & \xmark{} & \cmark{} & \xmark{} & \xmark{} & \cmark{} & \textbf{6.46} & \textbf{5.03} & \textbf{.658} & \textbf{2.10} & \textbf{3.91} \\
\midrule
C0 & \cmark{} & \cmark{} & \cmark{} & \xmark{} & \xmark{} & 3.15 & 4.65 & \textbf{.783} & \textbf{2.35} & 3.98 \\
C1 & \cmark{} & \cmark{} & \cmark{} & \cmark{} & \xmark{} & \underline{8.16} & \underline{4.33} & \underline{.780} & \underline{2.28} & 3.97 \\
C2 & \cmark{} & \cmark{} & \cmark{} & \xmark{} & \cmark{} & 4.95 & 4.37 & .757 & 2.19 & \textbf{4.14} \\
C3 & \cmark{} & \cmark{} & \cmark{} & \cmark{} & \cmark{} & \textbf{9.63} & \textbf{4.16} & .749 & 2.17 & \underline{4.03} \\
\bottomrule
\end{tabular}
\caption{Module ablation; S/A: SSL/acoustic encoders; $R$: predictor.}
\label{tab:ablation}
\end{table}

\paragraph{SSL-only path.}
A0 gives the highest P-ACC (12.45\%) but poor WER/SIM/PESQ, confirming that phoneme structure alone is not reconstruction sufficient. Direct adaptation in A1/A3 improves reconstruction but reduces P-ACC to 3.12\%/4.84\%; PAPA in A2/A4 restores it to 10.51\%/10.70\% while slightly lowering WER to 6.32/6.21.

\paragraph{Acoustic-only path.}
Without an SSL anchor, JMAS raises P-ACC from 4.60\% to 6.46\% and improves WER/SIM/PESQ/UTMOS from 5.27/.592/1.99/3.83 to 5.03/.658/2.10/3.91. It can therefore organize an acoustic bottleneck independently, but B1 remains below C3 on all five metrics.

\paragraph{Combined path.}
C0 attains the strongest SIM and PESQ in the C group but only 3.15\% P-ACC. PAPA (C1) and JMAS (C2) individually raise P-ACC and lower WER; C3 combines their gains to reach 9.63\% P-ACC and 4.16 WER. Relative to C1, C3 trades some SIM and PESQ for predictability and intelligibility, defining the intended operating point.

\begin{table}[!ht]
\centering
\small
\setlength{\tabcolsep}{0.5pt}
\begin{tabular}{@{}lcccccc@{}}
\toprule
Config. & Res. & CKA & KNN & Ret. & P-ACC & WER \\
\midrule
Direct (C0) & -- & 0.61 & 0.2627 & 40.62 & 3.15 & 4.65 \\
PAPA (C1) & 0.19 & 0.95 & 0.6340 & 98.02 & 8.16 & 4.33 \\
PAPA+JMAS (C3) & 0.20 & \textbf{0.96} & \textbf{0.6439} & \textbf{99.55} & \textbf{9.63} & \textbf{4.16} \\
\bottomrule
\end{tabular}
\caption{SSL-anchor retention; Res.: residual-to-anchor norm ratio; Ret.: KNN retention (\%).}
\label{tab:geometry-retention}
\end{table}

Table~\ref{tab:geometry-retention} quantifies geometric retention. Direct adaptation yields 40.62\%/0.61 KNN retention/CKA; PAPA restores 98.02\%/0.95, and JMAS reaches 99.55\%/0.96. The near-0.2 residual ratio suggests that the trainable branch does not overwrite the anchor. CKA uses centered raw frames without PCA, variance standardization, or $\ell_2$ normalization, supporting preservation rather than exact invariance.

\begin{figure}[t]
\centering
\includegraphics[width=\columnwidth]{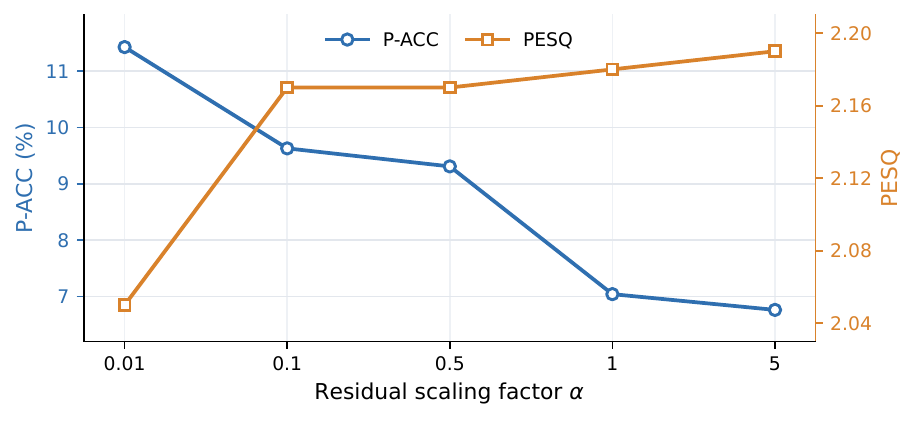}
\caption{PAPA residual-scale ablation.}
\label{fig:alpha-ablation}
\end{figure}

Figure~\ref{fig:alpha-ablation} shows that $\alpha=0.1$ balances 9.63\% P-ACC and 2.17 PESQ. Larger values improve PESQ only to 2.19 while lowering P-ACC to 6.76\%; $\alpha=0.01$ improves P-ACC but harms reconstruction. We use $\alpha=0.1$; Appendix D reports the other metrics.

\subsection{Downstream TTS}

\begin{table}[!ht]
\centering
\small
\setlength{\tabcolsep}{0.5pt}
\begin{tabular}{@{}lccccc@{}}
\toprule
Codec & Codebook & TPS & WER & SIM & UTMOS \\
\midrule
Ground Truth & -- & -- & 1.86 & .69 & 4.09 \\
WavTokenizer & 4096 & 75 & 23.10 & .38 & 4.01 \\
BigCodec & 8192 & 80 & 10.38 & \underline{.45} & \underline{4.21} \\
FocalCodec & 8192 & 50 & 6.82 & .44 & \underline{4.21} \\
X-Codec2 & 65536 & 50 & 9.49 & .43 & 4.15 \\
UniCodec & 16384 & 75 & 14.69 & .40 & 4.04 \\
AUV & 20480 & 50 & 12.98 & .44 & 4.19 \\
ReLMCodec (C1) & 8192 & 50 & 5.54 & \textbf{.48} & 4.14 \\
ReLMCodec (C3) & 8192 & 50 & \textbf{4.30} & \underline{.45} & \textbf{4.25} \\
ReLMCodec@64K & 65536 & 50 & \underline{4.93} & \textbf{.48} & \textbf{4.25} \\
\bottomrule
\end{tabular}
\caption{Downstream TTS; C1/C3 follow Table~\ref{tab:ablation}.}
\label{tab:tts}
\end{table}

\paragraph{Architecture benefit at matched token capacity.}
Table~\ref{tab:tts} compares all tokenizers under the same EmoVoice setup. ReLMCodec@64K and X-Codec2 each use 65{,}536 codewords at 50 TPS, enabling a capacity- and rate-matched architecture comparison. @64K improves WER/SIM/UTMOS from 9.49/.43/4.15 to 4.93/.48/4.25, supporting preserve--control--refine as a more effective autoregressive token interface.

\paragraph{Effect of JMAS.}
C1 and C3 both use an 8K, 50-TPS interface, isolating JMAS. C3 improves WER/UTMOS from 5.54/4.14 to 4.30/4.25 while SIM decreases from .48 to .45. Thus, teacher refinement strengthens linguistic consistency and naturalness with a modest speaker-similarity trade-off.

\paragraph{Codebook size versus downstream predictability.}
C3 and ReLMCodec@64K use identical architecture and a 50-TPS rate but 8K and 64K codebooks. The larger codebook restores SIM from .45 to .48, yet WER increases from 4.30 to 4.93 and UTMOS remains 4.25. Greater capacity preserves more speaker-specific variation but enlarges the autoregressive prediction space and reduces observations per token; it therefore does not automatically improve downstream performance. C3 better balances intelligibility and predictability; @64K favors speaker preservation.

\section{Limitations}

This study is limited to observational analysis of 50-Hz English LibriSpeech. PAPA requires aligned SSL/acoustic features, and teacher selection covers only W2v-BERT 2.0 and WavLM-Large L24; multilingual, noisy, and other-rate settings remain future work.

\section{Conclusion}

Our controlled probe shows that pre-quantization phoneme structure tracks autoregressive token predictability but is not sufficient for waveform reconstruction. ReLMCodec turns this diagnosis into a preserve--control--refine design: a frozen SSL anchor retains linguistic geometry, PAPA adds acoustic detail, and a training-only teacher refines post-quantization structure. At 650 and 800 bps, ReLMCodec improves this trade-off and downstream TTS.

\FloatBarrier
\appendix
\setcounter{table}{0}
\setcounter{figure}{0}
\setcounter{equation}{0}
\renewcommand{\thetable}{A\arabic{table}}
\renewcommand{\thefigure}{A\arabic{figure}}
\renewcommand{\theequation}{A\arabic{equation}}

\section{Appendix A: Analysis Reproducibility}

\subsection{Data Sampling and Phoneme Alignment}

The phoneme-structure analysis uses LibriSpeech~\citep{panayotov2015librispeech}. With a fixed random seed shared across representations, we sample 500 train-960 utterances as the reference set and 100 test utterances for phoneme analysis. This 500/100 sampling is used only to define the phoneme-analysis sets; token language-model probing uses LibriSpeech train-960 and the test splits.

\paragraph{Forced alignment.}
We resample LibriSpeech audio to 16 kHz and obtain phoneme boundaries using Montreal Forced Aligner (MFA) v3.3.9 with the English (US) ARPA acoustic model v3.0.0 and English (US) ARPA pronunciation dictionary v3.0.0. Utterances without a corresponding MFA-generated TextGrid are excluded from the phoneme analysis, and empty phone intervals are ignored. Stress markers are removed before ARPA vowels are merged into the final phoneme classes. The resulting boundaries are converted to frame-level labels at each encoder's native frame rate. The analysis covers 40 phoneme classes; to reduce class imbalance, at most 500 frames per phoneme are sampled from each split with the same fixed seed.

\subsection{Representation Extraction and Standardization}

Every encoder is frozen and evaluated without gradient updates. For codecs and tokenizers, we extract the continuous representation immediately before the native quantizer or discrete bottleneck; for self-supervised learning (SSL) models, we use the selected layer output named in Table~\ref{tab:appendix-baselines}. No temporal resampling is applied. For channel $j$, the train-960 statistics are
\begin{equation}
\mu_j = \frac{1}{N}\sum_{i=1}^{N} h_{i,j}, \qquad
\sigma_j = \sqrt{\frac{1}{N}\sum_{i=1}^{N}(h_{i,j}-\mu_j)^2 + \epsilon},
\end{equation}
and each frame is standardized as $\tilde{h}_{i,j}=(h_{i,j}-\mu_j)/\sigma_j$. Statistics are computed independently for every representation and are reused unchanged for its validation and test features.

\subsection{Phoneme Metrics and Visualization}

Principal component analysis (PCA) is fitted only on the 500-utterance reference set and reduces each standardized representation to 50 dimensions without whitening. K-nearest-neighbor (KNN) phoneme accuracy uses $k=10$ and is evaluated on the 100-utterance test set. Silhouette score, Davies--Bouldin index, and V-measure are computed on the same balanced frame samples.

The t-distributed stochastic neighbor embedding (t-SNE) panels are qualitative visualizations and do not enter the correlation analysis; all reported phoneme metrics follow the procedure above.

\subsection{Geometry Retention Metrics}

For the PAPA checkpoints C1 and C3, the relative residual magnitude gives every valid speech frame equal weight:
\begin{equation}
r_{\mathrm{res}}=\frac{1}{N}\sum_{i=1}^{N}
\frac{\|\alpha\Delta_i\|_2}{\|s_i\|_2+\epsilon}.
\end{equation}
It uses all valid frames from the 100 held-out utterances. The direct-adaptation checkpoint C0 has no anchor--residual decomposition, so its residual ratio is reported as ``--''. Centered kernel alignment (CKA) with a linear kernel instead uses exactly the phoneme-balanced held-out frame indices used by KNN: valid MFA-aligned frames from the same 100 utterances, 40 phoneme classes, and at most 500 frames per class. All configurations share utterances, frame indices, and random seed. For $S=[s_1,\ldots,s_N]^\top$ and $Z=[z_1,\ldots,z_N]^\top$, we center each feature column and compute
\begin{equation}
\operatorname{CKA}(S,Z)=
\frac{\|S^\top Z\|_F^2}
{\|S^\top S\|_F\,\|Z^\top Z\|_F}.
\end{equation}
CKA uses no variance standardization, PCA, or $\ell_2$ normalization. KNN retains the standardization, PCA, and reference/test procedure described above; KNN retention is $\mathrm{KNN}(z)/\mathrm{KNN}(s)$ with $\mathrm{KNN}(s)=0.6468$.

\subsection{Token--Phoneme Co-occurrence}

Token--phoneme co-occurrence uses the same MFA-aligned, phoneme-balanced held-out frames as the representation visualizations and KNN analysis. Each representation is discretized by its separately trained probing vector quantizer (P-VQ) under the shared 8K (8{,}192-codeword) configuration in Appendix B. For phoneme $p$ and token $k$, we count
\begin{equation}
C_{p,k}=\sum_i \mathbb{1}[y_i=p,\,k_i=k],
\end{equation}
and normalize each token column as $\bar{C}_{p,k}=C_{p,k}/\sum_{p'}C_{p',k}$. Token columns are grouped by the phoneme with maximum normalized co-occurrence. A concentrated column therefore indicates that the corresponding token is consistently associated with one phoneme, whereas diffuse off-block mass indicates sharing across phonemes. Appendix Figures~A3--A4 show all eight representations.

\section{Appendix B: Probing Quantizer and Token Language-Model Details}

\subsection{Probing Quantizer}

Native quantizers vary in codebook size, codebook count, quantizer family, and training objective, so native-token predictability conflates representation quality with quantizer capacity and design. We therefore bypass native quantizers where present and attach a matched probing quantizer to every representation. For a standardized representation $\tilde{h}\in\mathbb{R}^{d}$, the trainable input projection maps $d\rightarrow8$, the single-codebook P-VQ uses exponential moving average (EMA) updates to assign one of 8{,}192 codewords in $\mathbb{R}^{8}$, and the trainable output projection maps $8\rightarrow d$. Thus, representation dimensionality does not change the discrete capacity of the probe. The output projection reconstructs the continuous representation only; P-VQ contains no waveform decoder.

The probing objective is
\begin{equation}
\mathcal{L}_{\mathrm{probe}}
= \left\|P_{\mathrm{out}}(q)-\tilde{h}\right\|_2^2
+ \lambda_{\mathrm{commit}}\mathcal{L}_{\mathrm{commit}},
\end{equation}
with $\lambda_{\mathrm{commit}}=1$. Encoder features are detached, and only the projections and probing quantizer are optimized. All representations use the same data processing, number of updates, and quantizer configuration.

\begin{table}[!ht]
\centering
\small
\setlength{\tabcolsep}{3pt}
\begin{tabular}{lc}
\toprule
Setting & Value \\
\midrule
Codebook size & 8{,}192 \\
Codeword dimension & 8 \\
Input/output projections & $d\rightarrow8$ / $8\rightarrow d$ \\
Training steps & 200K \\
Segment duration & 3 seconds \\
Batch size & 16 \\
Reconstruction objective & $\ell_2$ feature reconstruction \\
Commitment weight & 1.0 \\
k-means initialization & 50 iterations \\
EMA update & Enabled; identical across probes \\
Inactive-code replacement & Enabled; identical across probes \\
Replacement source & Shared feature buffer \\
Optimizer and schedule & Identical across probes \\
Waveform decoder & None \\
\bottomrule
\end{tabular}
\caption{Shared probing-quantizer configuration.}
\label{tab:appendix-probe-config}
\end{table}

Inactive entries are replaced from a feature buffer populated by recent training batches. k-means initialization, EMA codebook updates, and inactive-code replacement are identical across all probes and prevent configuration differences from confounding the representation comparison.

\subsection{Autoregressive Token Language Model}

Each separately trained P-VQ produces a single token sequence at the representation's native frame rate. The same Qwen2-1.5B autoregressive model~\citep{yang2024qwen2} and training recipe are used for every representation. The model uses LibriSpeech train-960 and the test splits; the 500/100 phoneme-analysis samples do not define or alter these language-model splits.

\begin{table}[!ht]
\centering
\small
\setlength{\tabcolsep}{4pt}
\begin{tabular}{lc}
\toprule
Setting & Value \\
\midrule
Backbone & Qwen2-1.5B \\
Maximum sequence length & 2{,}048 \\
Learning rate & $10^{-4}$ \\
Warmup steps & 500 \\
Training epochs & 3 \\
Global batch size & 64 \\
Training split & LibriSpeech train-960 \\
Evaluation splits & LibriSpeech test-clean/test-other \\
Reported metrics & P-ACC and P-PPL \\
\bottomrule
\end{tabular}
\caption{Shared token language-model configuration.}
\label{tab:appendix-lm-config}
\end{table}

Probe-ACC (P-ACC) and Probe-PPL (P-PPL) are top-1 next-token accuracy and perplexity from P-VQ tokens under the shared Qwen2-1.5B recipe. P-PPL measures distribution-level modeling difficulty, while P-ACC provides an intuitive exact-prediction measure. Appendix Table~A5 instead reports native-token accuracy (N-ACC) for the two ReLMCodec reconstruction quantizers compared there; these metrics characterize differently optimized token spaces and are not directly comparable. Table~2 in the main text uses released checkpoints for baseline codecs and our trained checkpoints for ReLMCodec, while non-waveform representations in Appendix Figure~A1 use the matched-decoder diagnostic described in Appendix C. Codebook perplexity (CB-PPL) is $\exp(H(p(k)))$ under the assignments used by the corresponding row and is unrelated to P-PPL.

\section{Appendix C: Baselines and Reconstruction Protocols}

\paragraph{End-to-end codecs.}
All baseline codecs in Table~2 of the main text are evaluated using released complete checkpoints, including their original encoders, quantizers, and waveform decoders; no baseline component is retrained or replaced. ReLMCodec rows use our trained checkpoints. Reconstructions on test-clean and test-other are resampled to 16 kHz and scored with the same metric implementations.

The baseline checkpoints correspond to DAC~\citep{kumar2023high}, SpeechTokenizer~\citep{zhang2024speechtokenizer}, X-Codec~\citep{ye2025codec}, Stable Codec~\citep{parker2025scaling}, XY-Tokenizer~\citep{gong2026xy}, mimi~\citep{defossez2024moshi}, Qwen3-TTS-Tokenizer~\citep{hu2026qwen3}, UniCodec~\citep{jiang2025unicodec}, WavTokenizer~\citep{ji2025wavtokenizer}, X-Codec2~\citep{ye2025llasa}, AUV~\citep{chen2026auv}, SemantiCodec~\citep{liu2024semanticodec}, and FocalCodec~\citep{della2026focalcodec}.

The downstream text-to-speech (TTS) comparison additionally includes BigCodec~\citep{xin2024bigcodec}.

\paragraph{Matched-decoder diagnostics.}
The representation analysis also includes systems that are not complete waveform codecs. WavLM~\citep{chen2022wavlm}, HuBERT~\citep{hsu2021hubert}, and W2v-BERT~\citep{barrault2023seamless} release encoders only. MaskGCT~\citep{wang2025maskgct} includes an encoder, quantizer, and semantic decoder, but that decoder reconstructs continuous SSL features rather than waveforms. For reconstruction-based diagnostics such as SIM, we freeze the released representation modules and train the same waveform decoder with identical data, optimization, updates, and evaluation settings. These results measure representation reconstruction compatibility and appear only as diagnostic references in Appendix Figure~A1; they are not Table~2 codec baselines. Representation-level predictability separately uses P-VQ, except for Appendix Table~A5, which directly evaluates the two ReLMCodec quantizers.

\begin{figure}[!ht]
\centering
\includegraphics[width=\columnwidth]{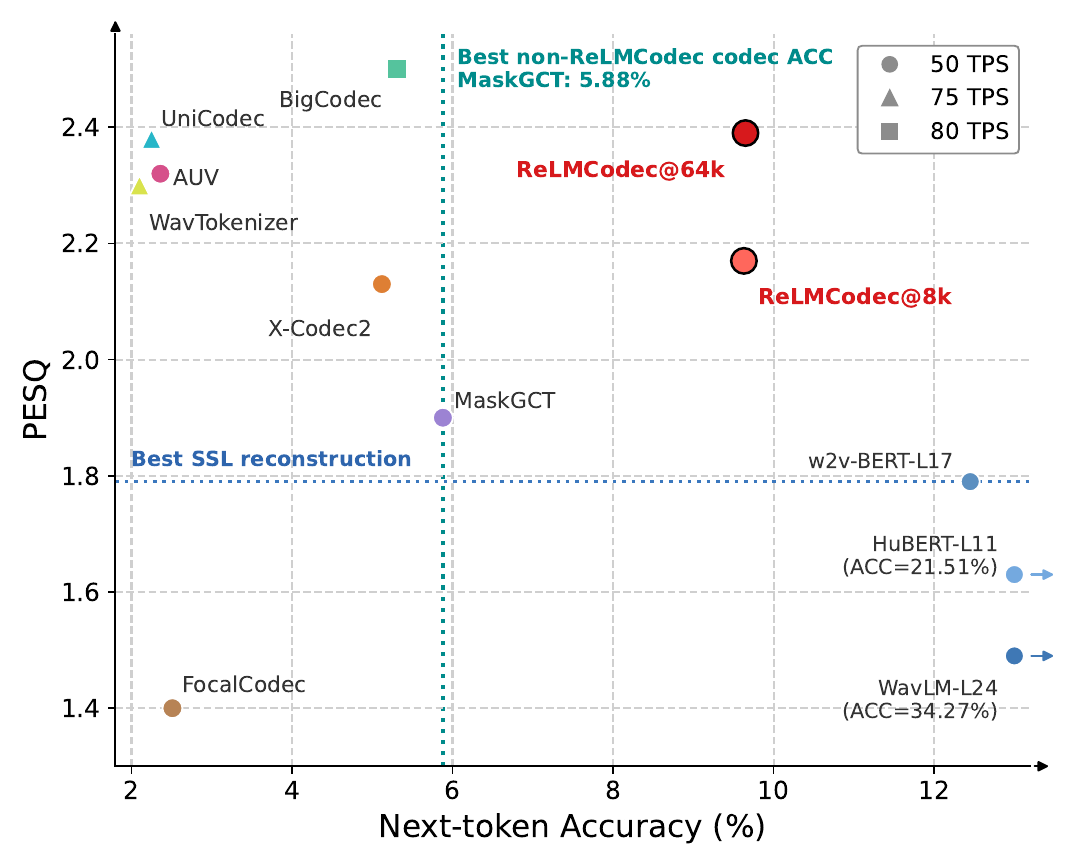}
\caption{P-ACC--PESQ diagnostic across waveform codecs and non-waveform references. The SSL and semantic-tokenizer references use matched waveform decoders and serve only as diagnostic context.}
\label{fig:appendix-frontier}
\end{figure}

\begin{figure*}[!b]
\centering
\begin{tabular}{@{}cc@{}}
\includegraphics[viewport=0 273.24 298.97 546.49,clip,width=0.29\textwidth]{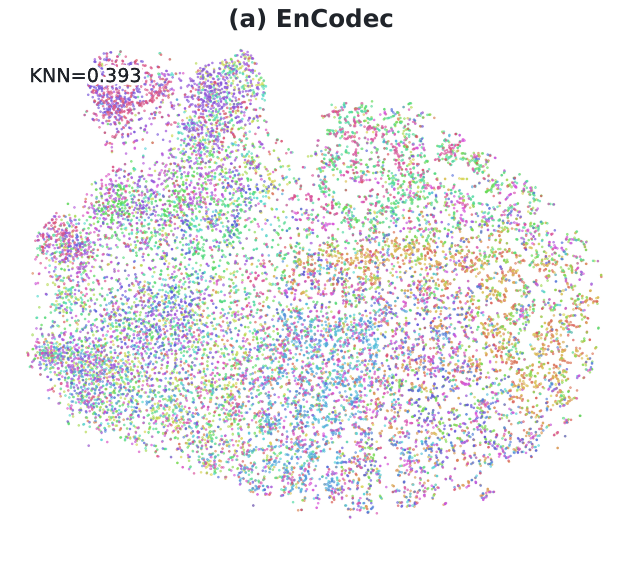} &
\includegraphics[viewport=298.97 273.24 597.95 546.49,clip,width=0.29\textwidth]{figures/tsne_pdfs/figure3_representative_tsne_phoneme_labeled_representations.pdf} \\
\includegraphics[viewport=597.95 273.24 896.92 546.49,clip,width=0.29\textwidth]{figures/tsne_pdfs/figure3_representative_tsne_phoneme_labeled_representations.pdf} &
\includegraphics[viewport=896.92 273.24 1195.89 546.49,clip,width=0.29\textwidth]{figures/tsne_pdfs/figure3_representative_tsne_phoneme_labeled_representations.pdf} \\
\includegraphics[viewport=0 0 298.97 273.24,clip,width=0.29\textwidth]{figures/tsne_pdfs/figure3_representative_tsne_phoneme_labeled_representations.pdf} &
\includegraphics[viewport=298.97 0 597.95 273.24,clip,width=0.29\textwidth]{figures/tsne_pdfs/figure3_representative_tsne_phoneme_labeled_representations.pdf} \\
\includegraphics[viewport=597.95 0 896.92 273.24,clip,width=0.29\textwidth]{figures/tsne_pdfs/figure3_representative_tsne_phoneme_labeled_representations.pdf} &
\includegraphics[viewport=896.92 0 1195.89 273.24,clip,width=0.29\textwidth]{figures/tsne_pdfs/figure3_representative_tsne_phoneme_labeled_representations.pdf}
\end{tabular}
\caption{Pre-quantization t-distributed stochastic neighbor embedding (t-SNE) representations for representative codec, tokenizer, ReLMCodec, and SSL features. Points are colored by phoneme label, and each panel reports KNN phoneme accuracy under the shared analysis protocol.}
\label{fig:appendix-representative-tsne}
\end{figure*}

The source papers for the additional Table~\ref{tab:appendix-baselines} rows are SoundStream~\citep{zeghidour2022soundstream}, HiFi-Codec~\citep{yang2023hificodec}, FunCodec~\citep{du2024funcodec}, Whisper~\citep{radford2022whisper}, MagiCodec~\citep{song2025magicodec}, LSCodec~\citep{guo2025lscodec}, UniAudio 1.5 and LLM-Codec~\citep{yang2024uniaudio}, PAST~\citep{hartuv2025past}, and SpeechTokenizer~\citep{zhang2024speechtokenizer}. Table labels that differ from paper titles are defined as follows: NeuCodec is the finite scalar quantization (FSQ) codec introduced by \citet{julian2025neucodec}; S3Tokenizer is the supervised semantic tokenizer introduced by CosyVoice~\citep{du2024cosyvoice}; and MingTok-Audio is the continuous tokenizer introduced by Ming-UniAudio~\citep{yan2025minguniaudio}. Discrete-WavLM6-KM uses the official \texttt{discrete\_wavlm\_large} release with \texttt{layer\_ids=[6]} associated with SELM~\citep{wang2024selm}, while SpeechTokenizer-Snake uses the official \texttt{speechtokenizer\_snake} checkpoint from SpeechTokenizer~\citep{zhang2024speechtokenizer}.

\subsection{Evaluation Metric Definitions}

\paragraph{Predictability and phoneme structure.}
P-ACC and P-PPL measure top-1 next-token accuracy and perplexity under the matched P-VQ protocol; higher P-ACC and lower P-PPL indicate easier autoregressive modeling. KNN accuracy, Silhouette, Davies--Bouldin, and V-measure quantify pre-quantization phoneme organization as detailed in Appendices A--B.

\paragraph{Reconstruction and downstream TTS.}
WER is computed with Whisper-Large-v3~\citep{radford2022whisper}. Multi-scale log-mel spectrogram loss (Log-Mel) compares reference and reconstructed speech at multiple time--frequency resolutions. SIM is the similarity between WavLM speaker embeddings~\citep{chen2022wavlm}; PESQ and STOI estimate perceptual quality and intelligibility, and UTMOS predicts speech naturalness~\citep{saeki2022utmos}. Lower WER and Log-Mel are better; higher SIM, PESQ, STOI, and UTMOS are better. The downstream TTS evaluation reports WER, SIM, and UTMOS under the shared generation setup.

\paragraph{Human evaluation.}
H-MOS is the human mean opinion score for reconstructed speech; higher ratings indicate better perceived quality. Appendix Figure~A5 reports rating distributions together with their medians and means.

\begin{figure*}[p]
\centering
\textbf{(a) EnCodec}\par
\includegraphics[width=0.90\textwidth]{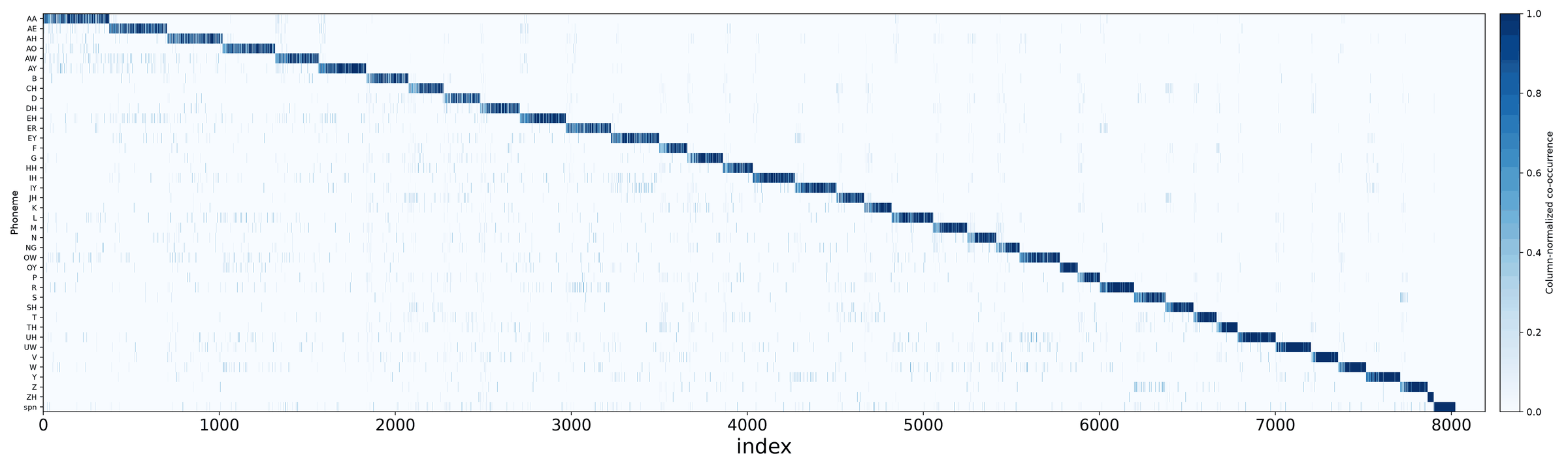}\par
\textbf{(b) WavTokenizer}\par
\includegraphics[width=0.90\textwidth]{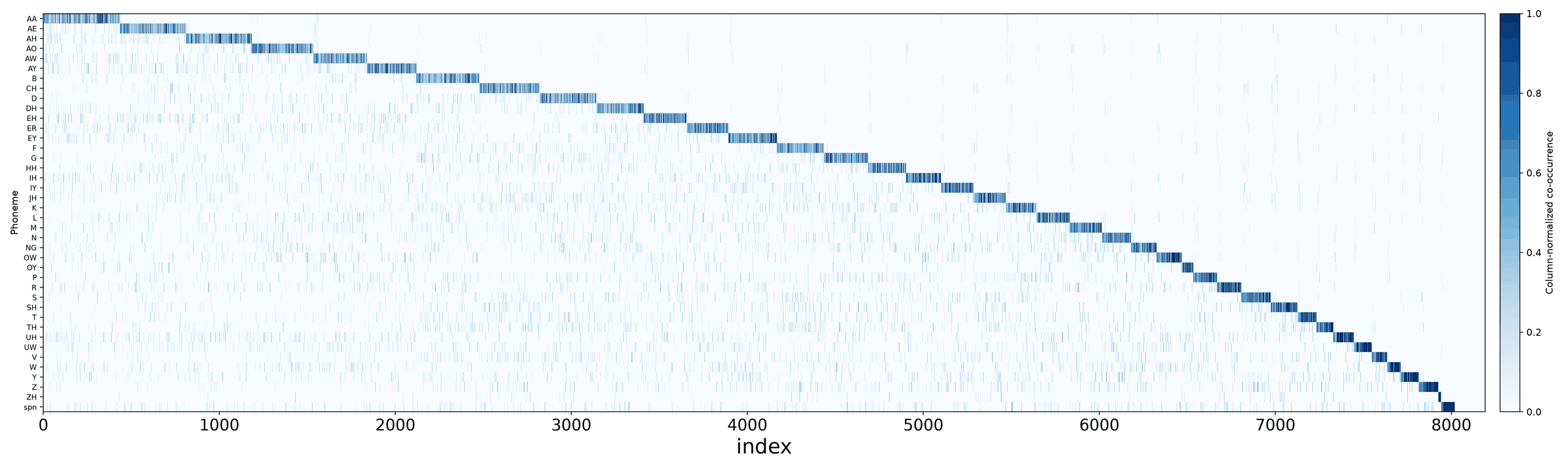}\par
\textbf{(c) BigCodec}\par
\includegraphics[width=0.90\textwidth]{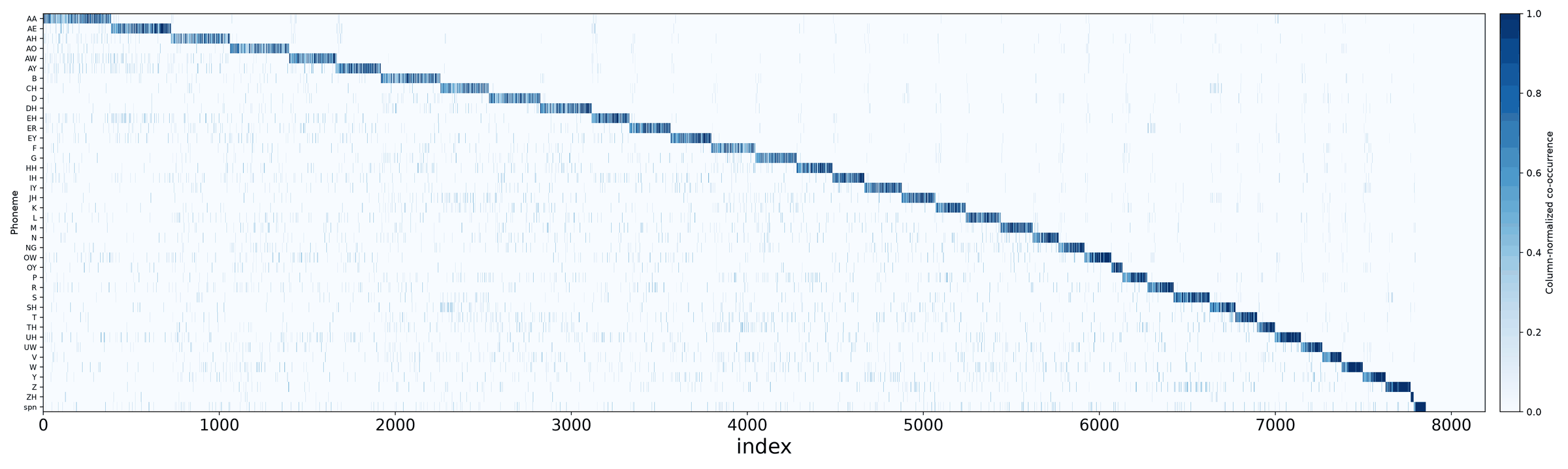}\par
\textbf{(d) X-Codec2}\par
\includegraphics[width=0.90\textwidth]{figures/paper_images/xcodec2_co_occurrence_heatmap.png}
\caption{Token--phoneme co-occurrence for representative waveform codecs under the matched 8K P-VQ protocol. Columns are normalized across phonemes and grouped by their dominant phoneme.}
\label{fig:appendix-token-phoneme-codecs}
\end{figure*}

\begin{figure*}[p]
\centering
\textbf{(a) MaskGCT}\par
\includegraphics[width=0.90\textwidth]{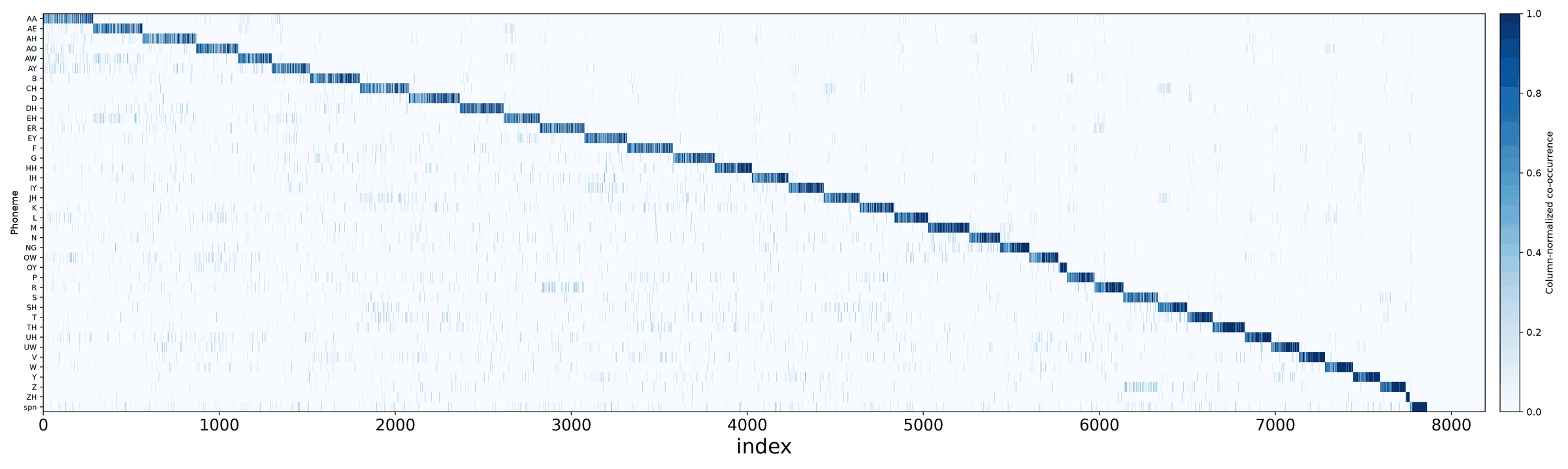}\par
\textbf{(b) ReLMCodec}\par
\includegraphics[width=0.90\textwidth]{figures/paper_images/relmcodec_co_occurrence_heatmap.png}\par
\textbf{(c) W2v-BERT 2.0 L17}\par
\includegraphics[width=0.90\textwidth]{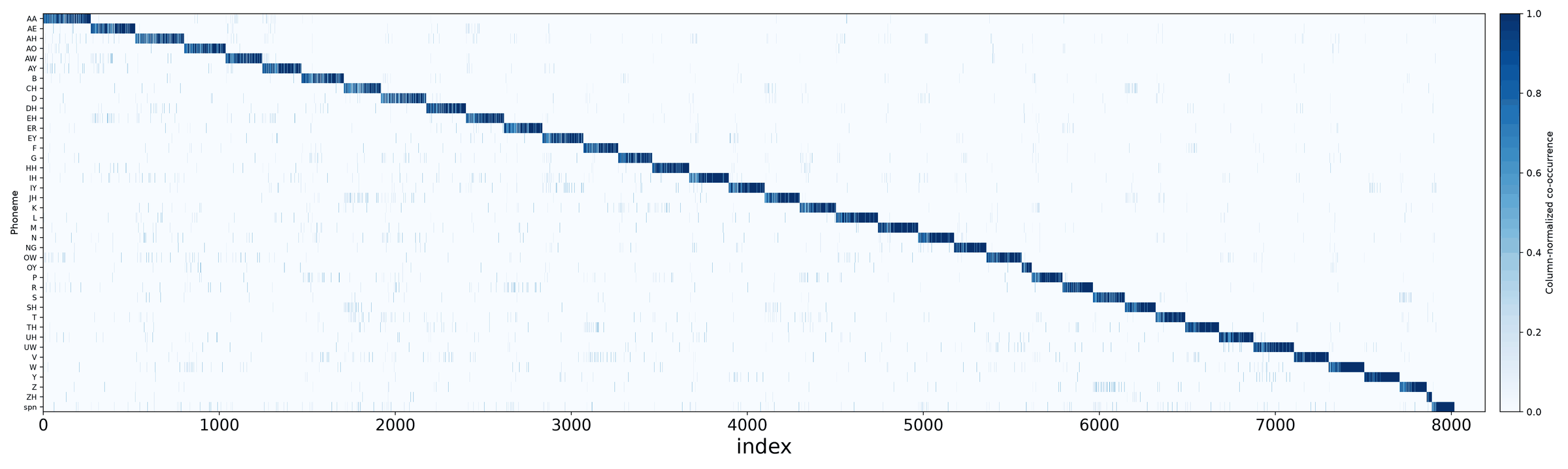}\par
\textbf{(d) WavLM-Large L24}\par
\includegraphics[width=0.90\textwidth]{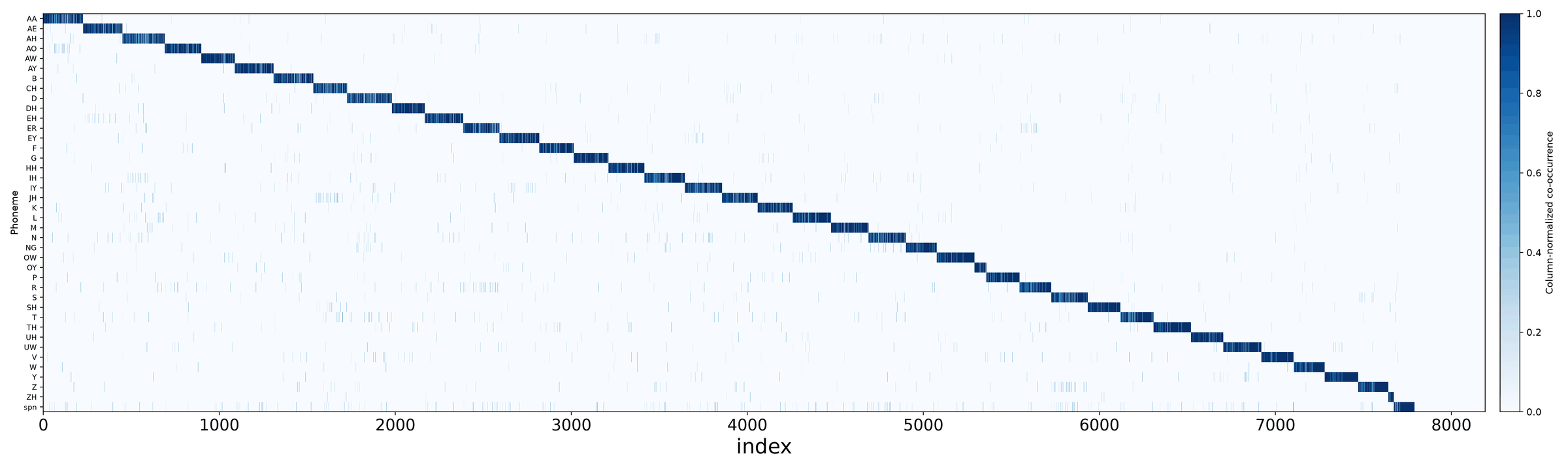}
\caption{Token--phoneme co-occurrence for MaskGCT, ReLMCodec, and SSL references under the same matched P-VQ protocol. ReLMCodec retains visibly concentrated phoneme--token structure relative to the representative waveform codecs in Appendix Figure~A3; the SSL panels provide reference organization.}
\label{fig:appendix-token-phoneme-ssl}
\end{figure*}

\FloatBarrier

\begin{table*}[t]
\centering
\small
\setlength{\tabcolsep}{1.2pt}
\begin{tabular}{@{}lrrrrrrr@{}}
\toprule
Model & TPS & P-PPL & P-ACC (\%) & KNN accuracy & Silhouette & Davies--Bouldin & V-measure \\
\midrule
SoundStream & 50 & 650.94 & 4.81 & 0.4560 & -0.0776 & 9.2685 & 0.2784 \\
HiFi-Codec & 50 & 274.65 & 5.42 & 0.3761 & -0.0894 & 9.8856 & 0.2400 \\
FunCodec & 50 & 278.27 & 5.56 & 0.4055 & -0.0958 & 10.4357 & 0.2428 \\
SpeechTokenizer & 50 & 150.88 & 9.85 & 0.7755 & 0.0131 & 4.3973 & 0.5041 \\
MaskGCT semantic codec & 50 & 228.35 & 5.88 & 0.4485 & -0.0870 & 8.5280 & 0.2592 \\
FocalCodec & 50 & 650.94 & 2.51 & 0.2970 & -0.0522 & 9.8502 & 0.1407 \\
X-Codec & 50 & 199.63 & 6.52 & 0.5140 & -0.0649 & 7.0121 & 0.3049 \\
X-Codec2 & 50 & 324.81 & 5.12 & 0.3374 & -0.0919 & 11.4711 & 0.1612 \\
AUV & 50 & 731.18 & 2.36 & 0.2969 & -0.0937 & 12.9768 & 0.1673 \\
SemantiCodec & 50 & 941.77 & 1.92 & 0.1947 & -0.0608 & 31.9050 & 0.0674 \\
W2v-BERT 2.0 L17 & 50 & 71.46 & 12.45 & 0.6468 & -0.0388 & 5.7575 & 0.3962 \\
WavLM-Large L24 & 50 & 11.33 & 34.27 & 0.7911 & 0.0419 & 3.5330 & 0.5159 \\
HuBERT-Large L11 & 50 & 21.41 & 21.51 & 0.8073 & 0.0121 & 3.8832 & 0.4946 \\
Whisper-Large-v3 L21 & 50 & 19.57 & 25.16 & 0.7883 & 0.0196 & 4.2970 & 0.4620 \\
NeuCodec & 50 & 351.62 & 4.86 & 0.3305 & -0.1008 & 13.4225 & 0.1619 \\
MagiCodec & 50 & 317.36 & 5.49 & 0.3864 & -0.0990 & 10.2949 & 0.2189 \\
LSCodec & 50 & 410.50 & 4.62 & 0.3245 & -0.1232 & 10.7016 & 0.1970 \\
SpeechTokenizer-Snake & 50 & 144.62 & 10.79 & 0.7974 & 0.0224 & 4.1380 & 0.5232 \\
LLM-Codec & 50 & 701.53 & 2.47 & 0.2996 & -0.1078 & 13.5889 & 0.1732 \\
UniAudio 1.5 & 50 & 286.86 & 5.51 & 0.3772 & -0.0317 & 12.2579 & 0.1878 \\
Discrete-WavLM6-KM codec & 50 & 65.84 & 13.41 & 0.7382 & -0.0114 & 4.5872 & 0.4641 \\
S3Tokenizer & 50 & 321.28 & 5.03 & 0.5423 & -0.0588 & 5.4132 & 0.3760 \\
PAST & 50 & 64.65 & 13.72 & 0.7176 & 0.0135 & 3.1359 & 0.5533 \\
MingTok-Audio & 50 & 334.57 & 5.01 & 0.3358 & -0.0585 & 13.5935 & 0.2046 \\
\bottomrule
\end{tabular}
\caption{Complete correlation set ($n=24$). TPS denotes tokens per second; all entries have native 50-Hz representations, yield 50-TPS P-VQ sequences, and are included in the main-paper Table~1 correlations. ReLMCodec is excluded from correlation computation.}
\label{tab:appendix-baselines}
\end{table*}

\begin{figure*}[t]
\centering
\includegraphics[width=0.72\textwidth]{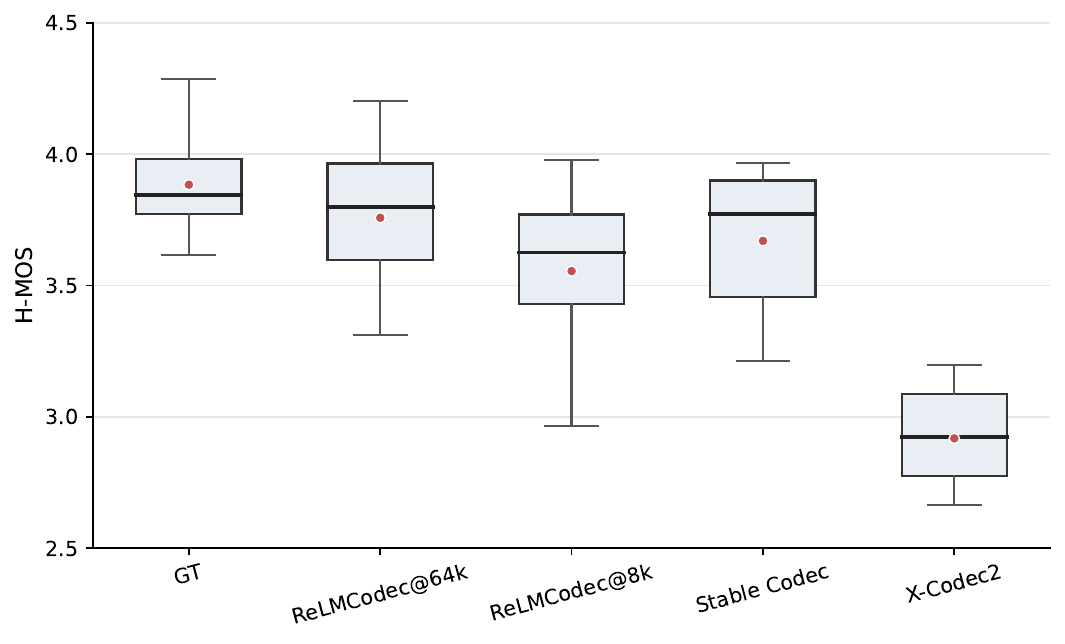}
\caption{Human mean opinion score (H-MOS) for reconstructed speech. ReLMCodec@64K has the highest mean H-MOS among the evaluated codecs and is closest to ground truth; the mean for ReLMCodec@8K also exceeds that of X-Codec2. Boxes summarize rating distributions, center lines show medians, and red markers show means.}
\label{fig:appendix-hmos}
\end{figure*}

\FloatBarrier

\section{Appendix D: Statistical and Ablation Details}

\subsection{Correlation Procedure}

Pearson and Spearman correlations between each separability metric and P-ACC/P-PPL use all $n=24$ rows of Table~\ref{tab:appendix-baselines}. Every row has a native frame rate of 50 Hz, so no listed model is removed by the frame-rate criterion. ReLMCodec is excluded. Confidence intervals use 10{,}000 row-wise bootstrap resamples, and $p$-values use 10{,}000 two-sided label permutations, both with random seed 2027. Main-paper Table~1 reports the rank correlations; Table~\ref{tab:appendix-correlations} gives complete statistics.

\begin{table}[!ht]
\centering
\footnotesize
\setlength{\tabcolsep}{2pt}
\begin{tabular}{@{}lcc@{}}
\toprule
Metric & $r$ vs. P-ACC & $\rho$ vs. P-ACC \\
\midrule
KNN & 0.806 [0.72, 0.93] & 0.911 [0.76, 0.98] \\
Silhouette & 0.787 [0.68, 0.89] & 0.715 [0.39, 0.88] \\
Davies--Bouldin & -0.574 [-0.83, -0.50] & -0.837 [-0.93, -0.61] \\
V-measure & 0.760 [0.68, 0.91] & 0.852 [0.67, 0.93] \\
\bottomrule
\end{tabular}

\smallskip

\begin{tabular}{@{}lcc@{}}
\toprule
Metric & $r$ vs. P-PPL & $\rho$ vs. P-PPL \\
\midrule
KNN & -0.812 [-0.89, -0.70] & -0.901 [-0.98, -0.73] \\
Silhouette & -0.614 [-0.80, -0.39] & -0.713 [-0.88, -0.40] \\
Davies--Bouldin & 0.804 [0.63, 0.93] & 0.848 [0.63, 0.94] \\
V-measure & -0.793 [-0.88, -0.67] & -0.857 [-0.94, -0.68] \\
\bottomrule
\end{tabular}
\caption{Pearson and Spearman correlations with 95\% bootstrap confidence intervals. All Spearman permutation $p$-values reported in main-paper Table~1 are below 0.001; the Pearson exceptions are Davies--Bouldin versus P-ACC ($p=0.015$) and Silhouette versus P-PPL ($p=0.002$).}
\label{tab:appendix-correlations}
\end{table}

\subsection{Residual-Scale Intervention}

Main-paper Figure~4 visualizes P-ACC and PESQ across the five residual scales. At $\alpha=5$, KNN/P-ACC/WER are 0.5847/6.76/4.11, and SIM/PESQ/UTMOS are 0.753/2.19/4.12. At $\alpha=1$, they are 0.6035/7.04/4.18 and 0.749/2.18/4.09, respectively. At $\alpha=0.5$, they are 0.6299/9.31/4.05 and 0.751/2.17/4.05. At $\alpha=0.1$, they are 0.6439/9.63/4.16 and 0.749/2.17/4.03. At $\alpha=0.01$, they are 0.6465/11.43/5.78 and 0.699/2.05/3.85. Every P-ACC uses a separately trained P-VQ rather than native codec token IDs. Across $\alpha>0.1$, PESQ changes by at most 0.02, while P-ACC falls to 6.76 at $\alpha=5$; decreasing $\alpha$ to 0.01 improves P-ACC but degrades WER, SIM, PESQ, and UTMOS. We therefore select $\alpha=0.1$ as the balanced operating point.

\subsection{Teacher-Selection Analysis}

With the W2v-BERT 2.0 L17 main path fixed, replacing the W2v-BERT teacher with WavLM-Large L24 improves P-ACC and WER, leaves SIM nearly unchanged, and decreases PESQ slightly. WavLM-Large L24 as the main path raises KNN and P-ACC but worsens WER, SIM, and PESQ. This two-model comparison supports the selected asymmetric assignment but does not establish robustness to arbitrary teacher architectures or domains.

\subsection{Quantizer Ablation}

We compare the native exponential-moving-average vector quantizer (EMA-VQ) with finite scalar quantization (FSQ).

\begin{table}[!ht]
\centering
\footnotesize
\setlength{\tabcolsep}{1.2pt}
\begin{tabular}{@{}lccccccc@{}}
\toprule
Quant. & N-ACC & WER & SIM & PESQ & UTMOS & Use. (\%) & CB-PPL \\
\midrule
EMA-VQ & 8.72 & 4.16 & 0.749 & 2.17 & 4.03 & 100.0 & 7445 \\
FSQ & 6.56 & 5.57 & 0.745 & 2.03 & 4.02 & 99.1 & 4961 \\
\bottomrule
\end{tabular}
\caption{Quantizer ablation for ReLMCodec C3 at 8K under identical settings. N-ACC uses native reconstruction tokens.}
\label{tab:appendix-vq-fsq}
\end{table}

EMA-VQ raises N-ACC from 6.56\% to 8.72\% and lowers WER from 5.57 to 4.16. Codebook usage corresponds to dead-code rates of 0\% and 0.9\%, while CB-PPL reaches 90.9\% and 60.6\% of nominal capacity for EMA-VQ and FSQ, respectively. Since P-VQ is trained for matched feature reconstruction whereas native quantizers use waveform objectives, P-ACC and N-ACC are not directly comparable; N-ACC is excluded from the correlation analysis.

\FloatBarrier

\section{Appendix E: Computational Cost and Inference Efficiency}

\subsection{Benchmark Protocol}

All models are evaluated in 32-bit floating point (FP32) with batch size 1 on 10-second, 16-kHz waveform inputs. We use one NVIDIA GeForce RTX 4090D GPU with 23.52 GiB memory and 15 allocated Intel Xeon Platinum 8474C CPU cores, running Ubuntu 22.04.5 LTS. The software environment uses Python 3.12.3, PyTorch 2.8.0 with CUDA 12.8, and cuDNN 9.1. Each measurement follows 10 warm-up iterations and 30 timed runs. Table~\ref{tab:appendix-efficiency} reports median real-time factor (RTF), with 10th--90th percentile (P10--P90) intervals in brackets, together with parameter counts and profiler-estimated floating-point operations (FLOPs). The accompanying benchmark metadata also records means, standard deviations, quartiles, and interquartile ranges for every stage.

The input waveform tensor is resident on the GPU before timing. RTF includes SSL feature extraction, model-internal preprocessing and transfers, PAPA, quantization, and waveform decoding where applicable. It excludes audio file I/O, waveform loading and resampling, the initial host-to-device transfer, and model loading. We call \texttt{torch.cuda.synchronize()} immediately before and after each timed invocation. Encode, decode, and end-to-end latency are measured independently; their medians are therefore not expected to be additive.

\begin{table}[!ht]
\centering
\footnotesize
\setlength{\tabcolsep}{2pt}
\begin{tabular}{@{}lcc@{}}
\toprule
Model & Params (M) & FLOPs / 1 s (G) \\
\midrule
ReLMCodec@8K & 782.74 & 61.74 \\
ReLMCodec@64K & 782.74 & 61.78 \\
X-Codec2 & 822.74 & 65.39 \\
\bottomrule
\end{tabular}

\smallskip

\begin{tabular}{@{}llc@{}}
\toprule
Model & Stage & Median [P10, P90] \\
\midrule
ReLMCodec@8K & Encode & 9.994 [9.986, 10.001] \\
 & Decode & 0.550 [0.548, 0.552] \\
 & End-to-end & 10.057 [9.987, 10.267] \\
ReLMCodec@64K & Encode & 9.987 [9.710, 10.005] \\
 & Decode & 0.555 [0.552, 0.557] \\
 & End-to-end & 10.324 [10.205, 10.967] \\
X-Codec2 & Encode & 9.993 [9.565, 9.999] \\
 & Decode & 0.935 [0.933, 0.942] \\
 & End-to-end & 10.070 [9.968, 10.148] \\
\bottomrule
\end{tabular}
\caption{Inference efficiency comparison. RTF values are reported as $\times10^3$ median [P10, P90] over 30 runs after 10 warm-up iterations. FLOPs are operator-counted by the PyTorch profiler for one second of audio; unsupported operations such as discrete lookup may be omitted.}
\label{tab:appendix-efficiency}
\end{table}

RTF interquartile ranges ($\times10^3$; encode/decode/end-to-end) are 0.0069/0.0020/0.1693 for ReLMCodec@8K, 0.1577/0.0031/0.2874 for ReLMCodec@64K, and 0.0088/0.0037/0.1009 for X-Codec2. ReLMCodec@8K matches X-Codec2 in median end-to-end RTF (10.057 versus 10.070), while ReLMCodec@64K is slightly slower (10.324). Both ReLMCodec variants use 4.86\% fewer parameters and approximately 5.5\% fewer operator-counted FLOPs. Their median decoding RTF is 41.17\% and 40.66\% lower for the 8K and 64K variants, respectively. Encoding latency is similar across the three systems and dominates end-to-end runtime. EMA codebooks are buffers, so both variants have identical parameter counts; their persistent states contain approximately 782.89M and 783.93M elements.

\paragraph{Code release.}
Upon acceptance, we will release the ReLMCodec checkpoints, probing scripts, representation statistics, and evaluation configuration.

\FloatBarrier
\bibliographystyle{plainnat}
\bibliography{references}

\end{document}